\documentclass[aps,prb,twocolumn,showpacs]{revtex4}
\usepackage{amssymb}
\usepackage{graphicx}
\usepackage{amsmath}
\usepackage{color}
\usepackage{bm}

\usepackage[normalem]{ulem}

\begin{document}
\newcommand {\beq} {\begin{equation}}
\newcommand {\eeq} {\end{equation}}
\newcommand {\bqa} {\begin{eqnarray}}
\newcommand {\eqa} {\end{eqnarray}}
\newcommand {\ba} {\ensuremath{b^\dagger}}
\newcommand {\Ma} {\ensuremath{M^\dagger}}
\newcommand {\psia} {\ensuremath{\psi^\dagger}}
\newcommand {\psita} {\ensuremath{\tilde{\psi}^\dagger}}
\newcommand{\lp} {\ensuremath{{\lambda '}}}
\newcommand{\A} {\ensuremath{{\bf A}}}
\newcommand{\Q} {\ensuremath{{\bf Q}}}
\newcommand{\kk} {\ensuremath{{\bf k}}}
\newcommand{\kp} {\ensuremath{{\bf k'}}}
\newcommand{\rr} {\ensuremath{{\bf r}}}
\newcommand{\rp} {\ensuremath{{\bf r'}}}
\newcommand {\ep} {\ensuremath{\epsilon}}
\newcommand{\nbr} {\ensuremath{\langle ij \rangle}}
\newcommand {\no} {\nonumber}
\newcommand{\up} {\ensuremath{\uparrow}}
\newcommand{\dn} {\ensuremath{\downarrow}}

\newcommand{\tb}[1]{\textcolor{magenta}{#1}}


\begin{abstract}

We study entanglement transitions in a periodically driven Ising
chain in the presence of an imaginary transverse field $\gamma$ as a
function of drive frequency $\omega_D$. In the high drive amplitude
and frequency regime, we find a critical value $\gamma=\gamma_c$
below which the steady state half-chain entanglement entropy,
$S_{L/2}$, scales with chain length $L$ as $S_{L/2} \sim \ln L$; in
contrast, for $\gamma>\gamma_c$, it becomes independent of $L$. In
the small $\gamma$ limit, we compute the coefficient, $\alpha$, of
the $\ln L$ term analytically using a Floquet perturbation theory
and trace its origin to the presence of Fisher-Hartwig jump
singularities in the correlation function of the driven chain. We
also study the frequency dependence of $\gamma_c$ and show that
$\gamma_c \to 0$ at special drive frequencies; at these frequencies,
which we analytically compute, $S_{L/2}$ remain independent of $L$
for all $\gamma$. This behavior can be traced to an approximate
emergent symmetry of the Floquet Hamiltonian at these drive
frequencies which we identify. Finally, we discus the behavior of
the driven system at low and intermediate drive frequencies. Our
analysis shows the presence of volume law behavior of the
entanglement in this regime $S_{\ell} \sim \ell$ for small subsystem
length $\ell \le \ell^{\ast}(\omega_D)$. We identify
$\ell^{\ast}(\omega_D)$ and tie its existence to the effective
long-range nature of the Floquet Hamiltonian of the driven chain for
small subsystem size. We discuss the applicability of our results to
other integrable non-hermitian models.

\end{abstract}
\title{Entanglement transitions in a periodically driven non-Hermitian Ising chain}

\author{Tista Banerjee and K. Sengupta }
 \affiliation{School of Physical Sciences, Indian Association for the Cultivation of
 Science, Jadavpur, Kolkata 700032, India.}

\date{\today}

\maketitle
\section{Introduction}
\label{intro}

Entanglement is a key feature of quantum many-body systems
\cite{entangrev1,entangrev2,cardy1,ryu1,hasting1,others1}. Its properties lead to
important information about nature of many-body quantum states. For
example, it is expected that a reduced density matrix $\rho_{\rm
red}$ constructed from the ground state (and the low-lying excited
states) of any $d$-dimensional gapped local quantum Hamiltonian
displays area law scaling: $S_{\ell} \sim \ell ^{d-1}$, where $\ell$
denotes the linear subsystem dimension \cite{hasting1}. For gapless
Hamiltonians, $S_{\ell} $ receives an additional logarithmic
corrections $S \sim \ell^{d-1} \ln \ell$. For $d=1$ where these
statements can be rigorously proved, $S_{\ell} \sim \ln \ell$ at a
critical point; the coefficient of the log term is given by the
central charge $c$ of the conformal field theory which describes the
critical point \cite{cftentang1}. In contrast, $\rho_{\rm
red}$ corresponding to mid-spectrum eigenstates of a generic
Hamiltonian usually shows volume law entanglement: $S_{\ell} \sim
\ell^d$.

The physics of closed quantum systems subjected to a periodic drive
has received tremendous attention in recent past
\cite{rev1,rev2,rev3,rev4,rev4a,rev5,rev5a,rev6,rev7,rev8}. Such systems are
typically described by their Floquet Hamiltonian $H_F$ which is
related to their evolution operator $U$ by $U(T,0) = {\mathcal T}_t
\exp[-i \int_0^T dt' H(t')/\hbar] = \exp[-i H_F T/\hbar]$. The
interest in such systems is partly due to the presence of
experimental platforms such as ultracold atoms in optical lattice
where such phenomena may be experimentally tested
\cite{exprev1,exp1,exp2,exp3,exp4}. Moreover these driven systems
allow one to explore several phenomena such as time crystalline
state of matter \cite{tc1,tc2,tc3}, dynamical freezing
\cite{df1,df2,df3,df4,df5}, dynamical localization
\cite{dl1,dl2,dl3}, generation of topologically non-trivial
Floquet phases \cite{topo1,topo2,topo3,topo4,topo5}, dynamical transitions
\cite{dt1,dt2,dt3}, prethermal
Floquet realization of quantum scars \cite{bm1,bm2,bm3} and Hilbert space
fragmentation \cite{ks1}; most of these phenomena have no
equilibrium analogue.

More recently, there has been a lot of interest in study of
non-Hermitian quantum systems \cite{nhrev,
nonhlit1,nonhlit2,nonhlit3,nonhlit4,nonhlit5,
nonhlit6,nonhlit7,nonhlit8,nonhlit9,nonhlit10,nonhlit11,nonhlit12,nonhlit13}.
This is due to the fact that such systems can now be engineered in
several experimental platforms \cite{nhexp1} and that they often
serve as models for studying open quantum systems. An explicit
example of the latter phenomena constitutes study of an Ising spin
chain in the presence of a measuring operator which measures $\hat
n_j = (1+\tau_j^z)/2$ (where $\tau_j^z$ denotes the usual Pauli
matrix representing the spin on site $j$ of the chain) with a rate
$\gamma$ and in the so-called no-click limit
\cite{dalibard1,daley1}; this leads to a complex magnetic field term
$\sim \gamma$ in the effective Hamiltonian of the spin chain.
Moreover, such systems often exhibit several physical properties
that have no Hermitian analogue; these include presence of
exceptional points in the spectrum \cite{ep1,ep2,ep3,ep4,ep5},
realization of skin effect \cite{skinherm1,skinherm2,skinherm3}
leading to novel bulk-boundary correspondence \cite{ep3,ep5}. More
recently, quantum dynamics of such systems have also been studied
\cite{nhdyn1,nhdyn2, nhdyn3,nhdyn4,nhdyn5}. In particular, the
presence of an emergent approximate symmetry in driven non-Hermitian
Ising chain has been pointed out recently \cite{tb1}.

In this work, we study the nature of entanglement in a driven
non-Hermitian Ising chain whose Hamiltonian is given by
\begin{eqnarray}
H = -J \sum_j \tau_j^x \tau_{j+1}^x - (h(t)+i \gamma/2)\sum_j \tau^z_j
\label{isingmodel}
\end{eqnarray}
where $J>0$ denotes the strength of the ferromagnetic interaction
term while $h(t)+i\gamma/2$ denotes the complex transverse field. The
drive is implemented by making the real part of the transverse field
time dependent via a given protocol. In what follows, we shall use
either the continuous drive protocol for which
\begin{eqnarray}
h(t) &=& h_1 + h_0 \cos \left(\omega_D t\right) \label{contprot}
\end{eqnarray}
or the square-pulse protocol which can be described by
\begin{eqnarray}
h(t) &=& h_1 +(-) h_0 \quad {\rm for} \quad t>(\le) T/2.
\label{sqprot}
\end{eqnarray}
Here $T= 2\pi/\omega_D$ is the drive period and $h_0$ denotes the
drive amplitude. In the rest of this work, we shall mostly be
interested in the subsystem size ($\ell$) dependence of the
steady-state entanglement entropy $S_{\ell}$ of such driven system
as a function of $\omega_D$ and $\gamma$. We note that whereas such
dependence has been studied for quench protocol
earlier\cite{nhdyn5}; however, it has not been analyzed for
periodically driven systems.

The central results that we obtain from such a study are as follows.
First, in the high drive frequency and amplitude regime, we find a
phase transition for entanglement at a critical $\gamma_c$. For
$\gamma \le \gamma_c$, the entanglement show a log scaling with
subsystem size. In particular, we study the half-chain entanglement
$S_{L/2} \equiv S_{\ell=L/2}$ which scales as $S_{L/2} \sim \ln L$
for $\gamma \le \gamma_c$. For $\gamma >\gamma_c$, $S_{L/2}$ is
constant and is independent of the subsystem size. The critical
transition value $\gamma_c$ depends on the drive frequency in a
non-monotonic manner and vanishes at special drive frequencies
$\omega_D=\omega_m^{\ast}$ where an approximate emergent symmetry
controls the behavior of the driven chain \cite{tb1}. We note that
the role of such an emergent symmetry behind entanglement
transitions in these systems has not been pointed out so far in the
literature.

Second, we analyze this phase diagram and provide an analytic,
albeit approximate, expression of the entanglement entropy in the
thermodynamic limit. Our analytical result for $S$, which matches
quite well with exact numerics, allows us to explain the lack of
volume law scaling of the entanglement which is normally expected
for driven Ising chains; moreover, it provides an expression of the
coefficient of the leading $\ln L$ term leading to an analytic
understanding of the phase diagram in the high drive amplitude
regime. Our analysis also shows that the contribution of the jump
singularities \cite{fisher1,fwref1,fwref2} in the correlation
functions of the driven chain to $S$ is key to its logarithmic
dependence on $\ell$; thus it identifies the Fisher-Hartwig
singularities of the correlation functions as the reason for the
entanglement transitions. To the best of our knowledge, this
connection was not identified in the literature before.

Third, we provide numerical study of the entanglement in the low and
intermediate frequency regime. Our analysis in this regime allows us
to identify a length scale $\ell^{\ast}(\omega_D)$ which diverges at
low-frequency. For $\ell \le \ell^{\ast}(\omega_D)$, the effective
Floquet Hamiltonian for the subsystem behaves as an effective
long-range Hamiltonian; consequently, the entanglement shows
volume-law scaling in this regime for small $\gamma$: $S_{\ell} \sim
\ell $. As $\ell$ increases and exceeds $\ell^{\ast}(\omega_D)$, $S$
crosses over from linear to logarithmic scaling. For large $\gamma$,
$S$ remains independent of $\ell$ for all $\ell$; the transition
between these two regimes occurs at a critical value of $\gamma$
similar to the high drive-frequency regime.

The plan of the rest of the paper is as follows. In Sec.\ \ref{fpt}
we present our analytical results; in Sec.\ \ref{hdrive}, we analyze
the driven chain to obtain the analytic, albeit perturbative,
Floquet Hamiltonians for different drive protocols and identify
their emergent symmetry. We also present expressions for the
correlation functions and analyze their features. This is followed
by Sec.\ \ref{anres} where we present analytic computation of the
half-chain entanglement entropy. Next, in Sec.\ \ref{numres}, we
present our numerical result for $S_{\ell}$; the phase diagram in
the high drive amplitude and drive frequency regime is presented in
Sec.\ \ref{phd} and compared with the analytical results obtained in
Sec.\ \ref{anres} while the low and intermediate drive frequency
regimes are discussed in Sec.\ \ref{lient}.  Finally, we discuss our
main results and their applicability to other integrable models and
conclude in Sec.\ \ref{diss}. Some additional details of the
calculations are presented in the appendices.

\section{Analysis of the driven chain}
\label{fpt}

In this section, we shall provide our analytical results which turns
out to be accurate in the high drive amplitude regime. The basic
properties of the Floquet Hamiltonian are discussed in Sec.\
\ref{hdrive} while the calculation of the entanglement entropy is
charted in Sec.\ \ref{anres}.

\subsection{Perturbative Floquet Hamiltonian} \label{hdrive}

The driven Hamiltonian of the Ising chain in the presence of complex
transverse field is given by Eq.\ \ref{isingmodel}. The drive
protocol used could either be continuous (Eq.\ \ref{contprot}) or
discrete (Eq.\ \ref{sqprot}). In what follows, we shall obtain
analytic expression for the Floquet Hamiltonian which will be useful
for computation of the correlation functions for the driven model.
The analysis of this section will closely follow Ref.\
\onlinecite{tb1}.

We begin by mapping Eq.\ \ref{isingmodel} into a system of free
two-components fermions; this is achieved by the well-known
Jordan-Wigner transformation given by


\begin{eqnarray}
\tau_j^{+(-)} &=& \left(\prod_{\ell=1}^{j-1} - \tau_{\ell}^z \right)
c_j^\dagger(c_j), \quad \tau_j^z= 2c_j^{\dagger} c_j -1 \label{jw}
\end{eqnarray}
where $c_j$ denotes annihilation operator of the fermions on site
$j$.
\begin{eqnarray}
    \hat{c}_j = \frac{1}{\sqrt{N}} \sum_{k\in\text{BZ}} e^{i\pi/4} e^{-ikj}\hat{c}_k
\end{eqnarray}
where ${\rm BZ}$ indicates the Brillouin zone $-\pi \le k \le \pi$.
Using Eq.\ \ref{jw}, and denoting the two component fermion field in
momentum space as $\psi_k= (c_k, c_{-k}^{\dagger})^T$, where $c_k$
annihilates a fermion with momentum $k$, Eq.\ \ref{isingmodel} can
be written as
\begin{eqnarray}
H &=& 2 \sum_{k \in\text{BZ}/2} \psi_k^{\dagger} H_k \psi_k \nonumber\\
H_k &=& \sigma_z (h(t) - \cos k +i\gamma/2) + (\sigma_+ \sin k + {\rm
h.c.}) \label{fermham}
\end{eqnarray}
where $\vec \sigma = (\sigma_x, \sigma_y, \sigma_z)$ are standard
Pauli matrices in particle-hole space of the fermions, $J$ and the
lattice spacing $a$ is set to unity, ${\rm BZ}/2$ denotes positive
half of the Brillouin zone, and $\sigma^{\pm} = \sigma_x \pm i
\sigma_y$.

The first-order contribution to the Floquet Hamiltonian can be
computed perturbatively using standard prescription of FPT
\cite{rev8,tb1}. Following the derivation sketched in App.\
\ref{appa}, one obtains $H_F^{(1),c(s)} = \frac{i \hbar}{T}
U_1^{c(s)}(T)$ where
\begin{eqnarray}
H_F^{(1),c(s)} &=& \sum_k \psi_k^{\dagger}[ S_{1k} \sigma_z + (
S_{2k}\sigma^+ + {\rm h.c.})]\psi_k \label{flham}\\
S_{1k} &=& 2 (h_1-\cos(k)+ i \gamma/2), \nonumber\\
S_{2k} &=&  2 f^{c}(T) \sin k \quad {\rm cosine \, protocol} \nonumber\\
&=&  2 f^{s}(T) \sin k \; e^{-i h_0 T/\hbar} \quad {\rm square\, pulse
\, protocol} \nonumber
\end{eqnarray}
where $f^{c(s)}(T)$ are given by
\begin{eqnarray}
f^c(T) &=& J_0\left(4h_0/(\hbar \omega_D) \right), \nonumber\\
f^s(T)&=& \frac{\hbar}{h_0T}\sin\left(\frac{h_0T}{\hbar}\right),
\label{flham1}
\end{eqnarray}
where $J_0$ denotes the zeroth order Bessel function. We note that
there are special drive frequencies at which
$f^{c,s}=0$ and $[H_F^{(1)}, \sigma_z]=0$ leading to an emergent
symmetry. These frequencies are given by $\omega_D= \omega_m^{\ast
c(s)}$ where
\begin{eqnarray}
\omega_m^{\ast c } &=& \frac{4 h_0}{\hbar \eta_m}, \quad
\omega_m^{\ast s}= \frac{2h_0}{m\hbar}  \label{spfr}
\end{eqnarray}
where $m$ is an integer and $\eta_m$ denotes the $m^{\rm th}$ zero
of $J_0$. This symmetry is approximate since it is violated by
higher-order terms of the Floquet Hamiltonian \cite{tb1};
nevertheless, it was shown in Ref.\ \onlinecite{tb1} that such an
approximate symmetry leaves its imprint on the dynamics of the
system. Here we shall show that such a symmetry shapes the character
of entanglement transitions of the driven non-Hermitian Ising chain.
We also note here that the inclusion of the second order terms in
$H_F$ do not change its matrix structure; it modifies $S_{1k}$ and
$S_{2k}$ as shown in App.\ \ref{appa}. In what follows, we shall use
this form of $H_F$ to compute second-order perturbative results for
the correlation functions and $S_{\ell}$.

The evolution operator $U(nT)$ for both protocols can be expressed
in terms of the eigenvalues $E_{k}^{a}$ and eigenvectors $|a;
k\rangle$ ( where $a=1,2$) of the Floquet Hamiltonian as
\begin{eqnarray}
U(nT) &=& \prod_k  \sum_{a=1, 2} e^{-i E_k^a nT} |a; k\rangle
\langle a; k|  \label{wavdef}
\end{eqnarray}
Here and in the rest of this section, we shall drop the indices
$c,s$ indicating the protocol to avoid clutter. For perturbative
Floquet Hamiltonian, these quantities can be analytically obtained and are given by
\begin{eqnarray}
E_k^a &=& (-1)^a \sqrt{ S_{1k}^2 + |S_{2k}|^2} = (-1)^a (\epsilon_k + i\Gamma_k) \nonumber\\
|a; k\rangle &=& \left( \begin{array}{c} n_{z k}^a \\ n_{x k}^a + i
n_{y k}^a
\end{array} \right), \quad n_{x k}^a = \frac{{\rm Re} [S_{2k}]}{{\mathcal N_{ak}}}
=\frac{q_{ak}}{{\mathcal N}_{a k}} \nonumber\\
n_{y k}^a &=& -\frac{{\rm Im} [S_{2k}]}{{\mathcal N_{ak}}}
=\frac{q'_{ak}}{{\mathcal N}_{a k}} \nonumber\\
n_{z k}^a &=& \frac{S_{1k} +(-1)^a  E_k}{{\mathcal N_{ak}}} =
\frac{p_{ak}}{{\mathcal N}_{ak}}, \nonumber\\
{\mathcal N}_{a k} &=& \sqrt{ \big| S_{1k} +(-1)^a E_k \big|^2
+ |S_{2k}|^2} \label{fleigen}
\end{eqnarray}
Note that $n_{yk}^a=0$ for the continuous protocol
for which $S_{2k}$ is real. The exact expressions for $E_k^a$ and
$|a;k\rangle$, however, needs to be computed numerically for the
continuous drive protocol. Its computation for the square-pulse
protocol can be carried out analytically as shown in App.\
\ref{appb}.

The normalized wavefunction of the driven chain starting from a
state $|\psi_0\rangle = \prod_k (u_{0k} + v_{0k} c_k^{\dagger}
c_{-k}^{\dagger})|0\rangle$ can be written, in terms of these
Floquet eigenvalues and eigenvectors as $|\psi(nT)\rangle = \prod_k
|\psi_k(nT)\rangle$ where
\begin{widetext}
\begin{eqnarray}
|\psi_k (nT)\rangle &=& (u_k(nT) + v_k (nT) c_k^{\dagger}
c_{-k}^{\dagger}) |0\rangle, \quad u_k(nT) = \frac{\sum_{a} e^{-i
E_k^a nT} p_{ak} (u_{0 k}p_{a k}^{\ast} + v_{0k} (q_{a k}^{\ast}+i
q_{a k}^{' \ast}))}{{\mathcal D}_k(nT)} \nonumber\\
v_k(nT) &=&  \frac{\sum_{a} e^{-i E_k^a nT} (q_{a k}-i q'_{ak})
(u_{0k}p_{a k}^{\ast} +
v_{0k} (q_{a k}^{\ast}+i q_{a k}^{' \ast}))}{{\mathcal D}_k(nT)}  \label{wavf1} \\
{\mathcal D}_k(nT) &=& [|\sum_{a} e^{-i E_k^a nT} p_{ak} (u_{0
k}p_{a k}^{\ast} + v_{0k} (q_{a k} ^{\ast}+i q_{a k}^{' \ast}))|^2
+ |\sum_{a} e^{-i E_k^a nT} (q_{a k}- i q'_{a k}) (u_{0k}p_{a
k}^{\ast} + v_{0k} (q_{a k}^{\ast}+i q_{a k}^{' \ast}))|^2]^{1/2}
\nonumber
\end{eqnarray}
\end{widetext}
and $|0\rangle$ denotes the fermion vacuum.

It is well-known that the computation of entanglement entropy for
integrable Ising chains begins with analysis of the correlation
functions of the model. We denote these correlation functions,
computed, at the end of $n$ drive cycles, as
\begin{eqnarray}
\Pi_{x k}(nT) &=&  \langle \psi_k(nT)| (c^\dagger_k c^\dagger_{-k} + {\rm
h.c.}) |\psi_k(nT)\rangle \nonumber\\
&=& 2 {\rm Re}(u_k^{\ast}(nT) v_k(nT)) \nonumber\\
\Pi_{z k} (nT) &=&  -i\langle
\psi_k(nT) |(c_{-k} c_{k} - c_k^\dagger c_{-k}^\dagger) |\psi_k(nT)\rangle \nonumber\\
&=& 2 {\rm Im }(u_k^{\ast}(nT) v_k(nT))  \label{corrdef}\\
\Pi_{y k}(nT) &=& \langle \psi_k(nT)| (2 c_k^{\dagger}c_k
-1)|\psi_k(nT)\rangle \nonumber\\
&=& 2|v_k(nT)|^2-1 \nonumber
\end{eqnarray}

\begin{figure}
\rotatebox{0}{\includegraphics*[width= 0.99 \linewidth]{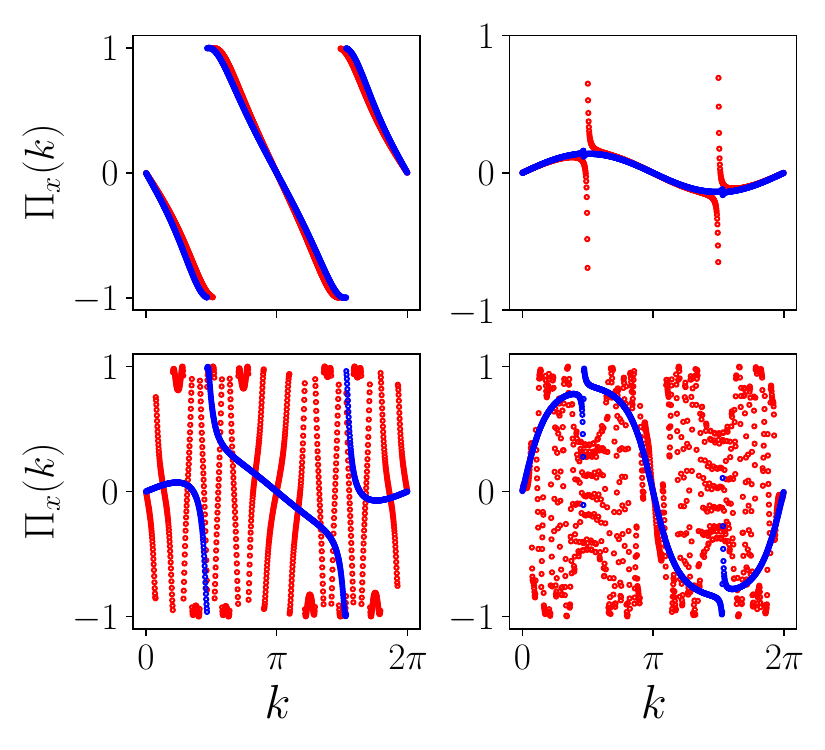}}
\caption{ Plot of $\Pi_{xk}$ as a function of $k$  for $\hbar
\omega_D= 60 J$ (top left panel), $\hbar \omega_D= \hbar
\omega_1^{\ast} \simeq 33.26 J$ (top right panel), $\hbar
\omega_D=0.7 J$ (bottom left panel) and $\hbar \omega_D=0.1 J$
(bottom right panel). The red lines display results from exact
numerics while the blue lines show that obtained using second order
Floquet perturbation theory. For all plots $\gamma=0.01 J$, $h_0=20
J$, $h_1=0.1 J$ and $L=1000$. See text for details. \label{fig1}}
\end{figure}

The plots of the steady state correlation function $\Pi_{xk}^{\rm
steady} \equiv \Pi_{xk}$ at large drive frequencies are shown in top
panels of Fig.\ \ref{fig1} for the continuous drive protocol and
$\gamma=0.01 J$. The coefficients of the steady state wavefunction
$|\psi_k^{\rm steady}\rangle$ are denoted by $u_k^{\rm steady}$ and
$v_k^{\rm steady}$ and are charted out in App.\ \ref{appb} for the
square pulse protocol. For the continuous protocol, the procedure is
analogous; however, the exact steady state wavefunction needs to be
obtained numerically. The steady state value of $\Pi_{x k}$ are
obtained by replacing $u_k(nT)$ and $v_k(nT)$ in Eq.\ \ref{corrdef}
by  $u_k^{\rm steady}$ and $v_k^{\rm steady}$ respectively. We have
checked numerically that $|\psi_k(nT)\rangle$ coincides with the
steady state correlation functions obtained using the
above-mentioned procedure after $n$ drive cycles, where $n$ depends on
both $\omega_D$ and $\gamma$, for both protocols. This feature has been
explicitly checked for all plots presented in this work.

From Fig.\ \ref{fig1}, we find that for generic drive frequencies
$\omega_D \ne \omega_m^{\ast}$, as shown in the top left panel of
Fig.\ref{fig1}, $\Pi_{x k}$ shows jump singularities at $k = \pm
k^{\ast}\simeq \arccos h_1$. In contrast, at the special drive
frequency $\omega_D= \omega_1^{\ast}$, the jump singularity
disappears and we obtain a smooth function for $\Pi_{xk}$. For
$\omega_D \ne \omega_m^{\ast}$, the second order Floquet result
(blue line) matches its exact numerical counterpart (red line) quite
well. In contrast, at $\omega_D=\omega_m^{\ast}$ and small values of
$\gamma$, one gets a qualitative match between the two results and
needs to go beyond 2nd order FPT for a quantitative match. At
intermediate ($\omega_D=0.7 J$) and low ($\omega_D=0.1
J$) drive frequencies, the perturbative Floquet theory clearly
breaks down as can be seen from the bottom panels of Fig.\
\ref{fig1}. Here exact numerics reveals multiple jump singularities
of $\Pi_{xk}$; the number of such jump singularities increases as
$\omega_D$ is lowered.

\begin{figure}
\rotatebox{0}{\includegraphics*[width= 0.99 \linewidth]{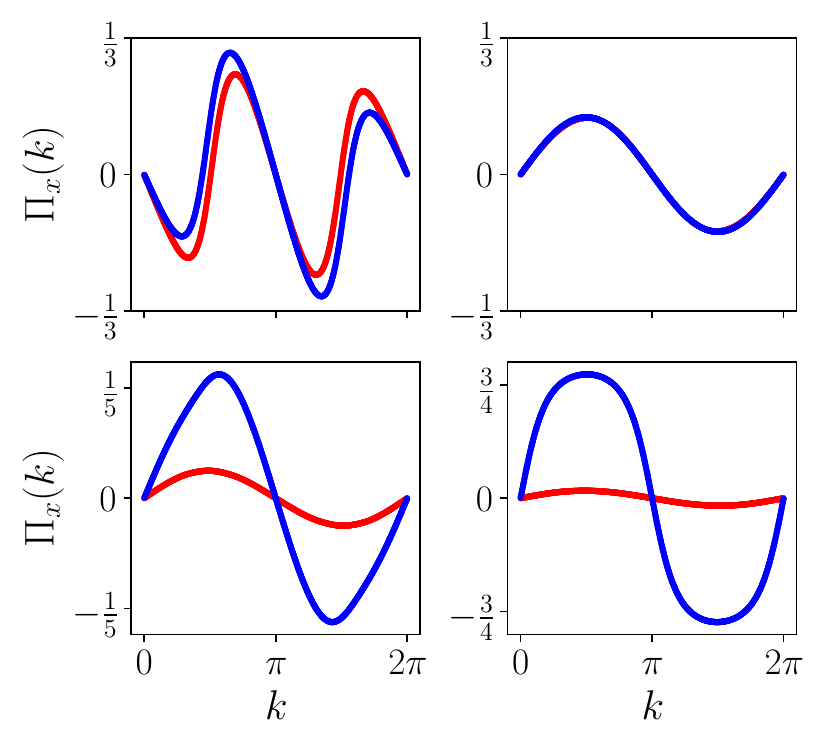}}
\caption{ Plot of $\Pi_{xk}$ as a function of $k$ for $\gamma=2J$.
All other parameters are same as in Fig.\ \ref{fig1}. See text for
details. \label{fig2}}
\end{figure}
In contrast at large $\gamma \sim J$, $\Pi_{xk}$ does not show any
jumps singularities. As shown in Fig.\ \ref{fig2} for $\gamma=2J$,
for all drive frequencies, $\Pi_{x k}$ is a smooth function of $k$.
The second order Floquet Hamiltonian, once again, yields a nice match
with exact numerics at the high drive frequency regime.

In contrast to $\Pi_{x k}$, the functions $\Pi_{yk}$ do not display
any jump singularities in the high drive frequency regime and stays
close to zero for $k\sim \pm k^\ast$. The behavior of $\Pi_{z k}$ is
qualitatively similar to $\Pi_{x k}$ except for the fact that the
magnitude of the jump singularity tends to zero at high frequencies
and low $\gamma/J$ where the contribution of $\Pi_{z k}$ to the
entanglement entropy becomes small. We shall use these features of
the correlation function to compute the entanglement entropy in the
next section.

\subsection{Analytical results for the entanglement entropy}
\label{anres}

In this section, we present an analytic computation of the
entanglement entropy $S_{\ell}$ for the high drive amplitude regime,
where the correlation functions obtained from perturbative Floquet
Hamiltomian agrees well with exact numerics. We shall carry out this
calculation in the regime where the dimensionless parameter $\gamma
\ll |\tilde {h}_1 f^{c(s)}(T)|\equiv |2\sqrt{1-h_1^2} f^{c,(s)}(T)|$.
Note that this indicates that such a computation is expected to yield
more quantitatively accurate results away from the special frequencies.

To compute $S_{\ell}$, we begin with expressions of $\hat{ \Pi}_k=
(\Pi_{xk}, \Pi_{yk}, \Pi_{zk})$ where $\hat {\Pi}_k$ denotes the
correlator values in the steady state. We first note that the
normalization of $u_k$ and $v_k$ leads to conservation of the norm
of the correlation functions $||\hat{\Pi}_k|| =1$. As is well-known
in the literature, under such conditions, Szeg\H{o}'s theorem
necessitates that $S_{\ell}$ can not have a term linear in $\ell$.
The derivation of this is sketched in App.\ \ref{appc}.

To obtain $S_{\ell}$, we first define the correlation matrix for the
system. To this end, we construct the quantity
\begin{eqnarray}
\eta_k &=& \lambda I -{\hat \Pi}_k \label{fdef}
\end{eqnarray}
which shall be central to computing $S_{\ell}$. In what follows we
shall provide an approximate analytic computation of $S_{\ell}$ in
the regime where $H_F^{(1)}$ provides a reasonable description of
the driven system. Following the calculations in App.\ \ref{appa}
and \ref{appb}, in this regime, one can define the steady state
wavefunctions to be
\begin{eqnarray}
|\psi\rangle &=& \prod_{k>0} \left(u_k^{(1)} + v_k^{(1)}\;
\hat{c}_k^\dagger \hat{c}_{-k}^\dagger|0\rangle \right) \nonumber\\
u_k^{(1)} &=& \frac{n_{z k} + \text{sgn}(\Gamma_k)}{{\mathcal N}_k}
\quad v_k^{(1)} = \frac{n_{xk} + in_{y k}}{{\mathcal N}_k}.
\label{steadyexp1}
\end{eqnarray}
where ${\mathcal N}_k= {\mathcal N}_{\pm k}$ for $\Gamma_k
>(<)0$. Substituting Eq.\ \ref{steadyexp1} in Eq.\ \ref{corrdef}, one obtain
the correlation functions $\hat \Pi^{(1)}_k$
\begin{eqnarray}
\Pi^{(1)}_{x k} &=& \frac{2\; {\rm Re}[(n_{z k} +
\text{sgn}(\Gamma_k))^\ast (n_{xk}+in_{yk})]}{{\mathcal N}_k^2} \nonumber\\
\Pi_{y k}^{(1)} &=& \frac{2\; {\rm Im}[(n_{z k} +
\text{sgn}(\Gamma_k))^\ast (n_{xk}+in_{yk})]}{{\mathcal N}_k^2}\nonumber\\
\Pi_{z k}^{(1)} &=& 1- 2\; \frac{|n_{zk} +
\text{sgn}(\Gamma_k)|^2}{{\mathcal N}_k^2}
\label{correxp1}
\end{eqnarray}
Near the jump singularities at $k= \pm k^{\ast}=\pm \arccos(h_1)$,
where $\Gamma_k$ changes sign, Eq.\ \ref{correxp1} can be further
simplified, for small $\gamma$, to yield
\begin{eqnarray}
\Pi^{(1)}_{x k } &=& \delta \; \text{Sgn}(\tilde{h}_1
f(T))\;\frac{k-{\rm Sgn}(k) k^\ast}{|k-{\rm Sgn}(k)k^\ast|}
\nonumber \\
\Pi^{(1)}_{z k } &=&  \sqrt{1-\delta^2} \;{\rm Sgn}(k), \quad \Pi_{y
k}^{(1)}=0 \nonumber \\
\delta &=& \sqrt{1-\left(\frac{\gamma}{\tilde{h}_1 f(T)}\right)^2}
\label{pires}
\end{eqnarray}
where we have ignored higher order terms in $k\pm k^{\ast}$. We note
that for small $\gamma$ and away from the special frequencies where
$|\delta| \sim 1$, $\Pi^{(1)}_{x k} \gg \Pi_{z k}^{(1)}$. In this
regime $\Pi_{x k}^{(1)}$ jumps around $k=\pm k^{\ast}$; these jumps
control the $\ell$ dependence of $S_{\ell}$, as shall be elaborated
later in this section. In contrast, for large $\gamma/J$ or around
$\omega_D= \omega_m^{\ast}$ at which $f(T) \simeq 0$, the
off-diagonal terms of the Floquet Hamiltonian will be small in the
high-drive frequency regime. This leads to a steady state which
closely mimics one of the eigenstates of $\hat{\tau}^z$ depending on
$\text{Sgn}\left[\Gamma_k\right]$. Hence $\Pi_{zk}^{(1)}$ dominates
at these frequencies and $\hat \Pi^{(1)}_k$ becomes a smooth
function of $k$ leading to a different behavior of $S_{\ell}$. The
transition between these behaviors which constitutes the
entanglement transition occurs around $\delta \sim 0$; thus $\delta$
controls the behavior of $S_{\ell}$ in this regime.

To obtain an analytic expression of $S_{\ell}$ for large $\ell$ we
now carry out a Fisher-Hartwig analysis
\cite{fisher1,fwref1,fwref2}. We note that  this analysis can be
done only when the correlation matrix depends on a single Pauli
matrix. The reason for this has been discussed extensively in the
literature \cite{fwref1,fwref2} and stems from the fact that
such an analysis requires the correlation matrix of the model to be in
the standard Toeplitz form \cite{fisher1}; for $\hat \Pi_{k}$ with multiple
components leading to block-Toeplitz form of the correlation matrix such an
analysis does not hold \cite{fwref1,fwref2}. We shall therefore focus on the regime
$\delta \sim 1$ where $\hat \Pi_k \simeq (\Pi_{x k},0,0)$ and focus
on the form of $\Pi_{x k}$ near $k=\pm k^{\ast}$ given by Eq.\
\ref{pires}. In this case, one can define
\begin{eqnarray}
\zeta_k &=& \lambda - \Pi_{x k}  \label{geneq1}
\end{eqnarray}
which acts as generators for the elements of the correlation matrix.

Next, we cast $\zeta_k$ in a form which is convenient for analysis
of contribution of the singularities. To this end, we note that the
form of $\Pi_{x k}$ given in Eq.\ \ref{pires} holds only near the
jump singularities at $k= \pm k^{\ast}$; the functional dependence
of $\Pi_{x k}$ over the entire Brillouin zone $-\pi \le k \le \pi$
is not captured by Eq.\ \ref{pires}. It is well-known that the
precise nature of this functional form is not important for
computing the contribution of the singularities to $S_{\ell}$
\cite{fisher1}. Thus one can replace $\Pi_{x k}$ away from the
singularities by an appropriate form which allows for further
analytical progress. The simplest such form will be to replace
$\Pi_{x k}=\delta$ everywhere away from the singularities at $k=\pm
k^{\ast}$. However since the jumps for $k= k^{\ast}$ and
$k=-k^{\ast}$ occur in opposite direction as one traverse from
$k=-\pi $ to $k=\pi$, one can not replace $\Pi_{x k}$ by a constant
without introducing intermediate spurious singularities, say, at
$k=0$ and $k=\pi$. Keeping this in mind, we use the form
\begin{eqnarray}
\Pi_{x k}^{(1)} &=& \text{Sgn}(\tilde{h}_1 f(T)) \; {\rm Sgn}(k-k^{\ast}), \quad 0< k <\pi \nonumber\\
&=&  \text{Sgn}(\tilde{h}_1f(T)) \;{\rm Sgn}(k + k^{\ast}), \:
-\pi< k <0 \label{pikform}
\end{eqnarray}
We shall use this form to compute $S_{\ell}$ analytically; the
additional contribution due to the spurious singularities shall be
subtracted out at the end of the calculation.

Using Eqs.\ \ref{geneq1} and \ref{pikform}, one can compute the
eigenvalues of the correlation matrix in the steady state
\begin{eqnarray}
{\mathcal D}_{\ell}(\lambda) &=& \lambda I - B_{\ell} \otimes
\sigma_x,\nonumber\\
B_{\ell} &=& \left ( \begin{array}{cccc} \Pi_0 & \Pi_{-1} & ... & \Pi_{1-\ell} \\
\Pi_1 & \Pi_{0} & ... & \Pi_{2-\ell} \\
... & ... & ... & ... \\ \Pi_{\ell-1} & \Pi_{\ell-2} & ... & \Pi_{0}
\end{array} \right), \nonumber\\
\Pi_{\ell} &=& \frac{1}{2\pi}\int_0^{2 \pi} dk \;e^{i \ell k} \;
\Pi_{x k}^{(1)} \label{corrmat1}
\end{eqnarray}
This calculation, which constitutes standard steps in Fisher-Hartwig
analysis \cite{fisher1, fwref1,fwref2} is charted out in App.\
\ref{appd} and yields
\begin{eqnarray}
\ln\left( \text{Det}\left( \mathcal{D}_{\ell}\left(\lambda\right)
\right) \right) &=& 2 [\ell \; \ln\left( F\left[\eta(\lambda)\right]
\right) - \sum_{i=1, 3} \beta_i^2(\lambda)\ln(\ell) ] +... \nonumber\\
F[\eta(\lambda)] &=& \sqrt{\lambda^2-\delta^2} \nonumber\\
\beta_1 &=& \beta_{3}= \frac{1}{2\pi i} \ln \left(\frac{\lambda-
\delta}{\lambda+\delta}\right) \label{dexp1}
\end{eqnarray}
where the ellipsis represent subleading terms which we shall ignore
for the rest of the analysis, we have subtracted out the
contributions from the spurious singularities at $k=0, \pi$, and
$\beta_{1,3}$ depicts the contribution of the jump singularities at
$\pm k^{\ast}$ to $\text{Det} \left({\mathcal D}_{\ell}\right)$.
Note that the first term in Eq.\ \ref{dexp1} constitutes
contribution from the non-singular part of $\eta_k$ as shown in
App.\ \ref{appd}. The contribution of such term to $S_{\ell}$
vanishes as shown in App.\ \ref{appc} for $\delta=1$. In the rest of
this section, we shall estimate the contribution of the second term
which yields the $\ln \ell$ behavior.

The contribution of the second term in the expression of $\ln
{\mathcal D}_{\ell}$ to $S_{\ell}$ can be computed by defining the
function $e(x,\lambda)$
\begin{equation}
e(x,\lambda) \equiv -\frac{x+\lambda}{2}
\ln{\frac{x+\lambda}{2}}-\frac{x-\lambda}{2}
\ln{\frac{x-\lambda}{2}} \label{ldef}
\end{equation}
The entanglement entropy can be written in the form $S_{\ell} =
\sum_{m=-\ell}^{\ell} e(\delta,\nu_m) $ where $\nu_m$ denote the
eigenvalues of the correlation matrix $B_{\ell}$ which lie within
the range $-\delta \le \nu_m \le \delta$ with the property $\nu_{m}=
-\nu_{-m}$. To connect $S_{\ell}$ to ${\mathcal D}_{\ell}$, we note,
from Eq.\ \ref{corrmat1}, that $ {\mathcal D}_{\ell}(\Lambda) =
\prod_{m=-\ell}^{\ell} (\lambda -\nu_m)$. Thus one can write
\begin{eqnarray}
S_{\ell} &=& \frac{1}{4 \pi i} \oint d \lambda \, e(\delta,
\lambda)\, \frac{d}{d\lambda} \ln \left( {\rm Det} \left({\mathcal
D}_{\ell}(\lambda)\right) \right) \label{sexp1}
\end{eqnarray}
where the contour encircles the zeroes of $\text{Det}
\left({\mathcal D}_{\ell}(\Lambda)\right)$. We now substitute Eq.\
\ref{dexp1} in Eq.\ \ref{sexp1} to evaluate $S_{\ell}$ and
concentrate on the coefficient of $\ln \ell$ in the limit of small
$\delta$. A straightforward calculation yields
\begin{eqnarray}
S_{\ell} &=& \delta I_1 \ln \ell + {\rm O}(\delta-1) \label{entrexp} \\
I_1 &=& \frac{2}{\pi^2} \oint d\lambda \;e(1,\lambda)
\frac{\beta_1(\lambda)}{\lambda^2-1} = \frac{2}{\pi^2}\int_0^1 dx
\;\frac{\ln(1-x)}{x} = \frac{1}{3} \nonumber
\end{eqnarray}
For $\delta=1$ which is achieved in the quench limit when $T, \gamma
\to 0$, we obtain $S_{\ell} =1/3 \ln \ell$ which
reproduces the result obtained in Ref.\ \onlinecite{nhdyn5} as a
special case. For the driven system, $\delta$ is a non-monotonic
function of the drive frequency whose precise form depends on the
drive protocol through $f(T)$. This allows us to infer that
$S_{\ell}(T)$ will be a non-monotonic function of $T$. We shall
compare this analytic form of $S_{\ell}$ with exact numerical
results at small $\gamma$ and high drive frequency regime in the
next section.

Before ending this section, we would like to discuss two salient
points of our analysis. First, the approximating $\hat \Pi_k$ with
only one of its components, namely $\Pi_{x k}$, violates the
normalization of $\hat \Pi(k)$ for all $\delta(T) \ne 1$. This in
turn leads to a spurious volume law term; this term can be obtained
when the first term of $\ln D_{\ell}(\lambda)$ (which is $\sim
\ell$) in Eq.\ \ref{dexp1} is substituted in Eq.\ \ref{sexp1}.
Second, the ${\rm O}(\delta-1)$ terms in Eq.\ \ref{entrexp} can not
be reliably computed within this approach; it leads to a logarithmic
divergence for $\delta \ne 1$. Both these features are artifacts of
violation of the normalization condition of $\hat \Pi_k$. A more
rigorous calculation keeping the full matrix structure of $\hat
\Pi_k$ which requires the use of Fredholm techniques may resolve
these issues; however, such a technique, used for computing
$S_{\ell}$ for ground state of the Hermitian XY model \cite{fwref2},
can not be straightforwardly applied in the present case due to
presence of a non-zero $\Gamma_k$. We leave this issue as subject of
future work.

\begin{figure}
\rotatebox{0}{\includegraphics*[width= 0.49 \linewidth]{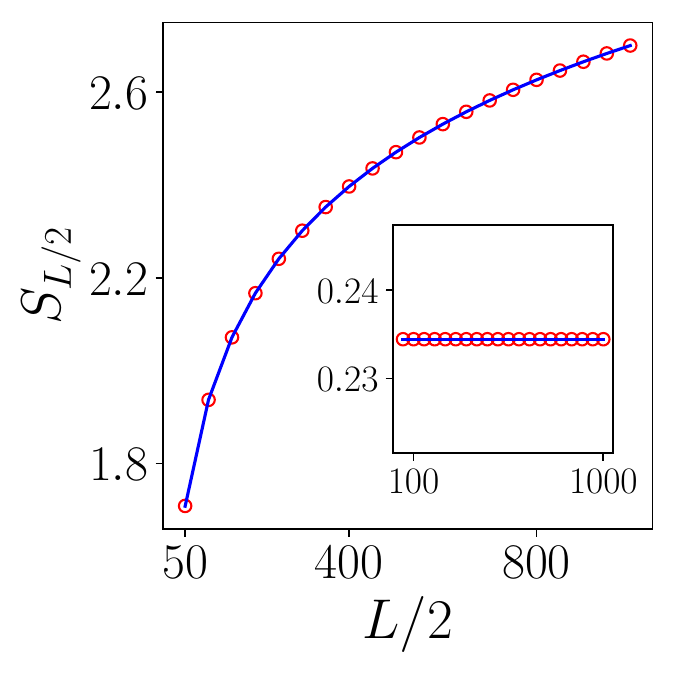}}
\rotatebox{0}{\includegraphics*[width= 0.49 \linewidth]{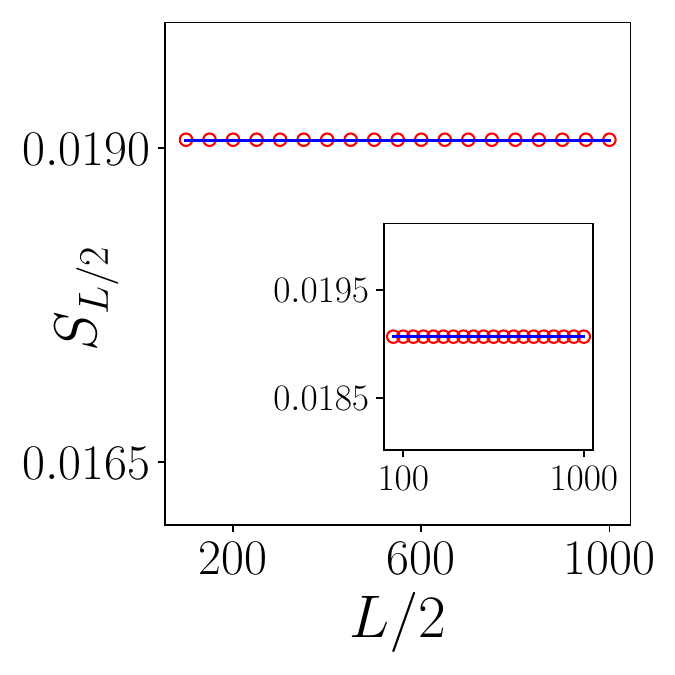}}
\caption{Left Panel: Plot of steady-state half-chain entanglement
$S_{L/2}$ as a function of $L$ showing logarithmic dependence of
$S_{L/2}$ at small $\gamma= 0.1J$ and $\hbar \omega_D/J=60$:
$S_{L/2} \sim \alpha \ln L  + {\rm constant}$. A fit of $S_{L/2}$
estimates $\alpha \sim 0.3314$ which is close to its analytically
predicted value $1/3$. The inset shows $S_{L/2}$ at $\gamma =2J$
indicating its independence on $L$: $S_{L/2} = {\rm constant}$.
Right Panel: Same as in left panel but at $\omega_D= \omega_1^{\ast
c}$ where $S_{L/2}$ is independent of $L$ for both small
($\gamma=0.1J$) and large ($\gamma=J$) $\gamma$. For all plots, the
red dots indicate numerical data and blue lines represent the fit,
$h_0=20 J$, $h_1=0.1 J$. See text for details. \label{fig3}}
\end{figure}

\section{Numerical results} \label{numres}

In this section, we present our numerical results on steady state
entanglement entropy $S_{\ell}$. The numerical procedure for
computing $S_{\ell}(nT)$ for the driven state is as follows. First,
for the continuous protocol, we carry out Trotter decomposition of
the evolution operator $U_k(T,0)$ and write
\begin{eqnarray}
U_k(T,0) &=& \prod_{j=1}^N U_k(t_j, t_{j-1})=  \prod_{j=1}^N e^{-i
H(t_j) \Delta t/\hbar} \label{ueq1}
\end{eqnarray}
where the interval $\Delta t= t_j-t_{j-1} = T/N$; it is chosen to be
small enough so that $H(t)$ does not change significantly within
this interval. For the square pulse protocol, $U_k(T,0)$ can be
exactly obtained as shown in App.\ \ref{appb}. For either case,
diagonalization of $U_k(T)$ leads to its eigenvalues $e^{\pm i E_k^F
T/\hbar}$, where $E_k^F$ are the exact Floquet eigenvalues, and the
corresponding eigenfucntions  are $|\pm; k \rangle$. This allows one
to write
\begin{eqnarray}
U_k(T,0) &=& \sum_{a=\pm} e^{-i a E_k^F T/\hbar} |a; k\rangle
\nonumber\\
|\psi_{k}( nT)\rangle &=& \frac{|\tilde \psi_{ k}(nT)\rangle}{|\langle \tilde \psi_{ k}(nT)|\tilde \psi_{ k}(nT)\rangle|},\nonumber\\
|\tilde \psi_{ k}(nT)\rangle &=&  U_{ k}(nT,0) |\psi_{0 k}\rangle
\label{exwav1}
\end{eqnarray}
Note that for non-Hermitian systems the non-conservation of norm
during evolution necessitates normalization of the wavefunction at
all stroboscopic times $nT$. Having obtained $|\psi_k(nT)\rangle$,
we use it to construct the correlation functions $\Pi_k(nT)$ using
Eq.\ \ref{corrdef} and follow standard procedure charted out earlier
to obtain $S_{\ell}(nT)$. At large $n$, $S_{\ell}(nT)$ reaches its
steady state value $S_{\ell}$ which we analyze below in details.

\begin{figure}
\rotatebox{0}{\includegraphics*[width= 0.48 \linewidth]{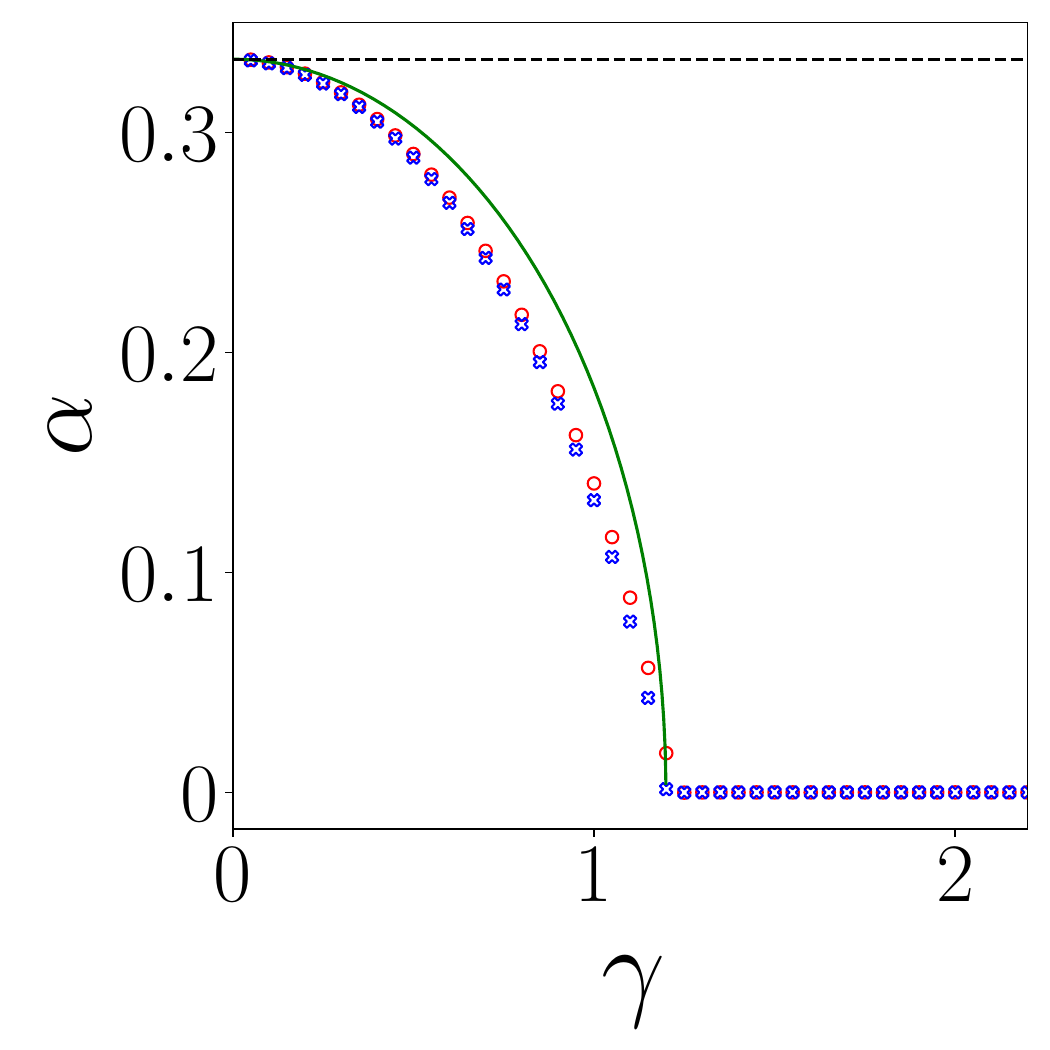}}
\rotatebox{0}{\includegraphics*[width= 0.48 \linewidth]{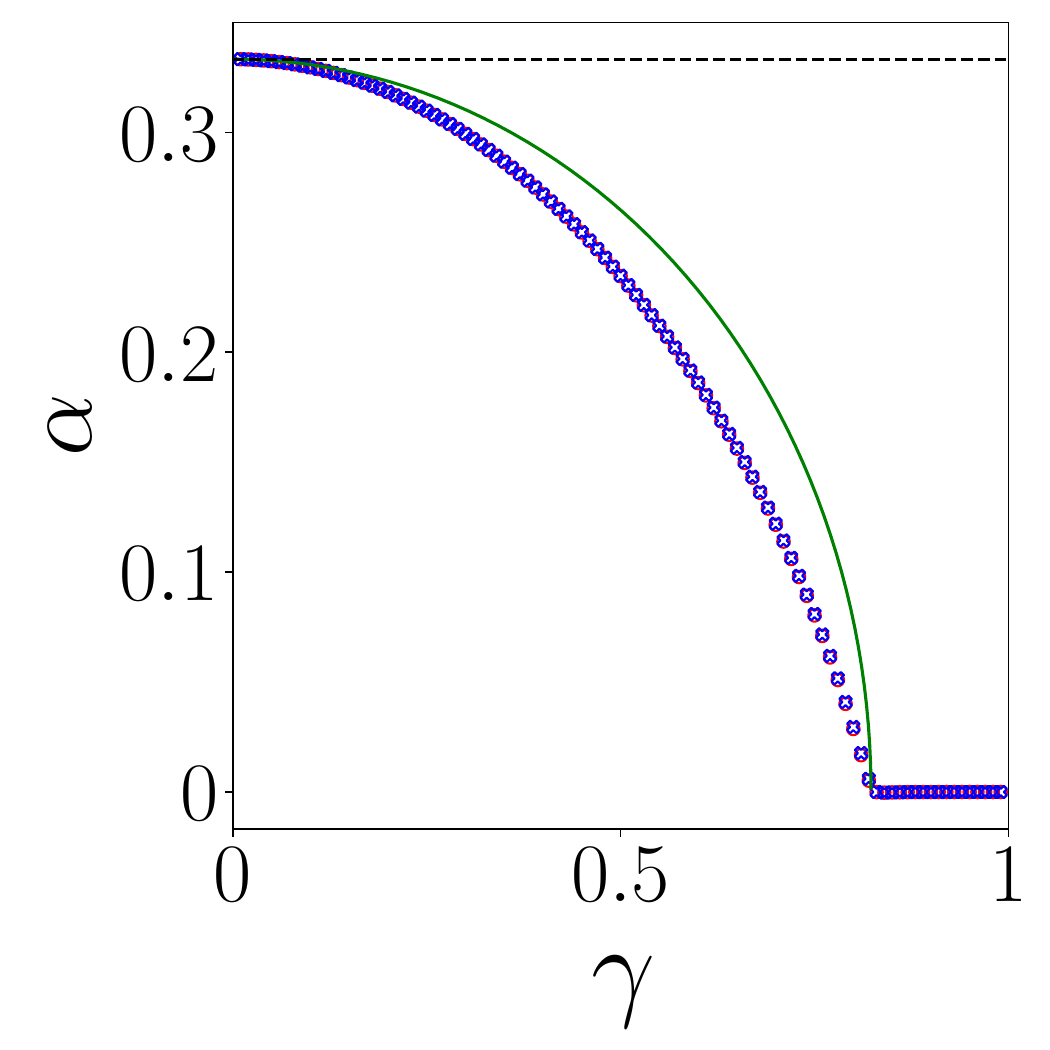}}
\caption{Left Panel: Plot of $\alpha$ as a function of $\gamma$ for
$\hbar \omega_D/J=60$ using continuous drive protocol, showing its
decrease from $1/3$ to $0$ at the critical $\gamma=\gamma_c \sim 1J$
where the entanglement transition takes place. The red and the blue
dots correspond to exact numerical results and results from second
order FPT respectively; the green line shows analytical result for
$\alpha$ obtained in Eq.\ \ref{entrexp}. Right panel: Same as left
panel using square pulse drive protocol. The red and the blue dots
correspond to exact numerical results and results from first order
FPT respectively while the green line shows analytical result for
$\alpha$ obtained in Eq.\ \ref{entrexp}. The dashed line in both
figures indicate $\alpha=1/3$ and is a guide to the eye. For all
plots $h_0=20 J$, $h_1=0.1 J$ and $L \le 2000$.}\label{fig4}
\end{figure}

\begin{figure}
\rotatebox{0}{\includegraphics*[width= 0.48 \linewidth]{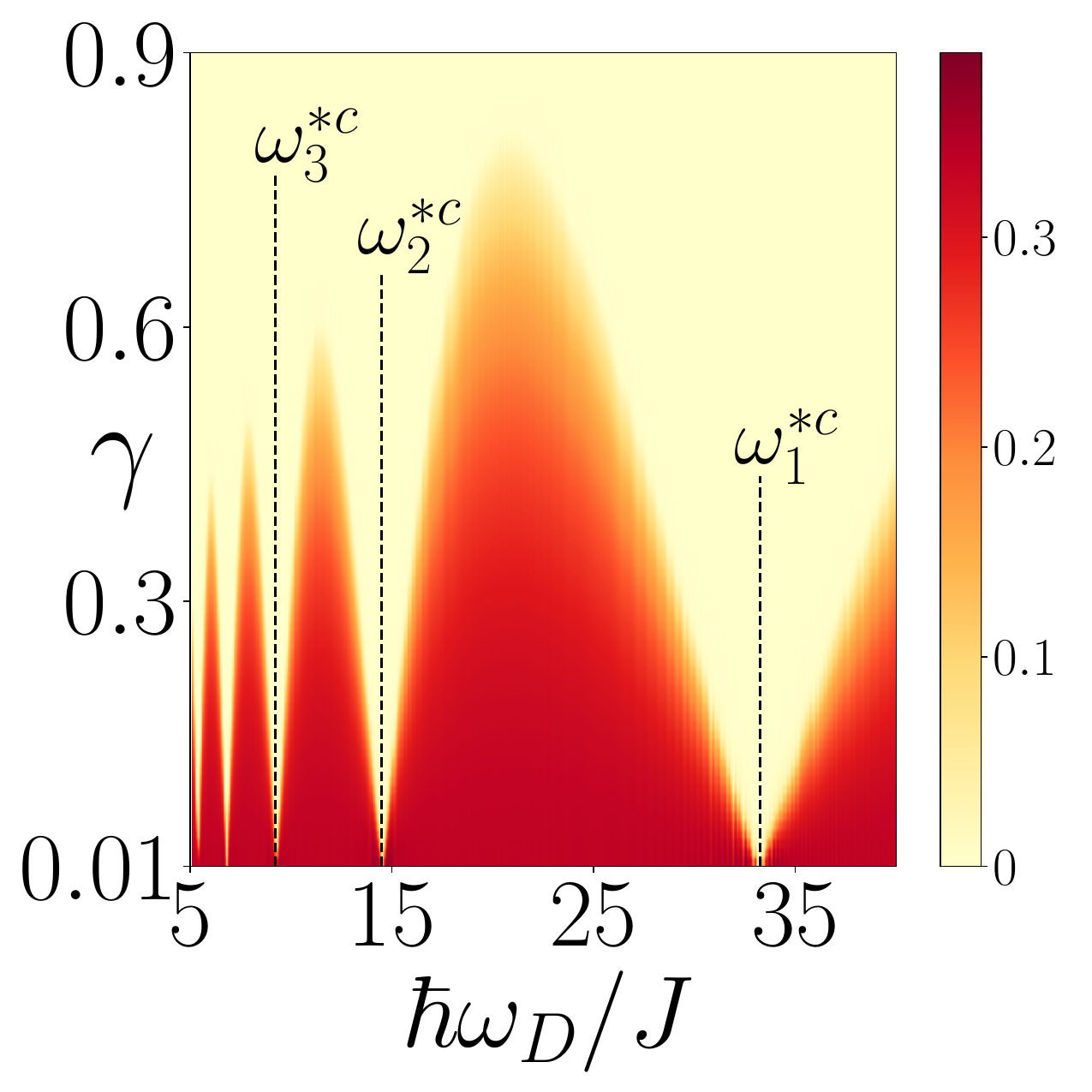}}
\rotatebox{0}{\includegraphics*[width= 0.48 \linewidth]{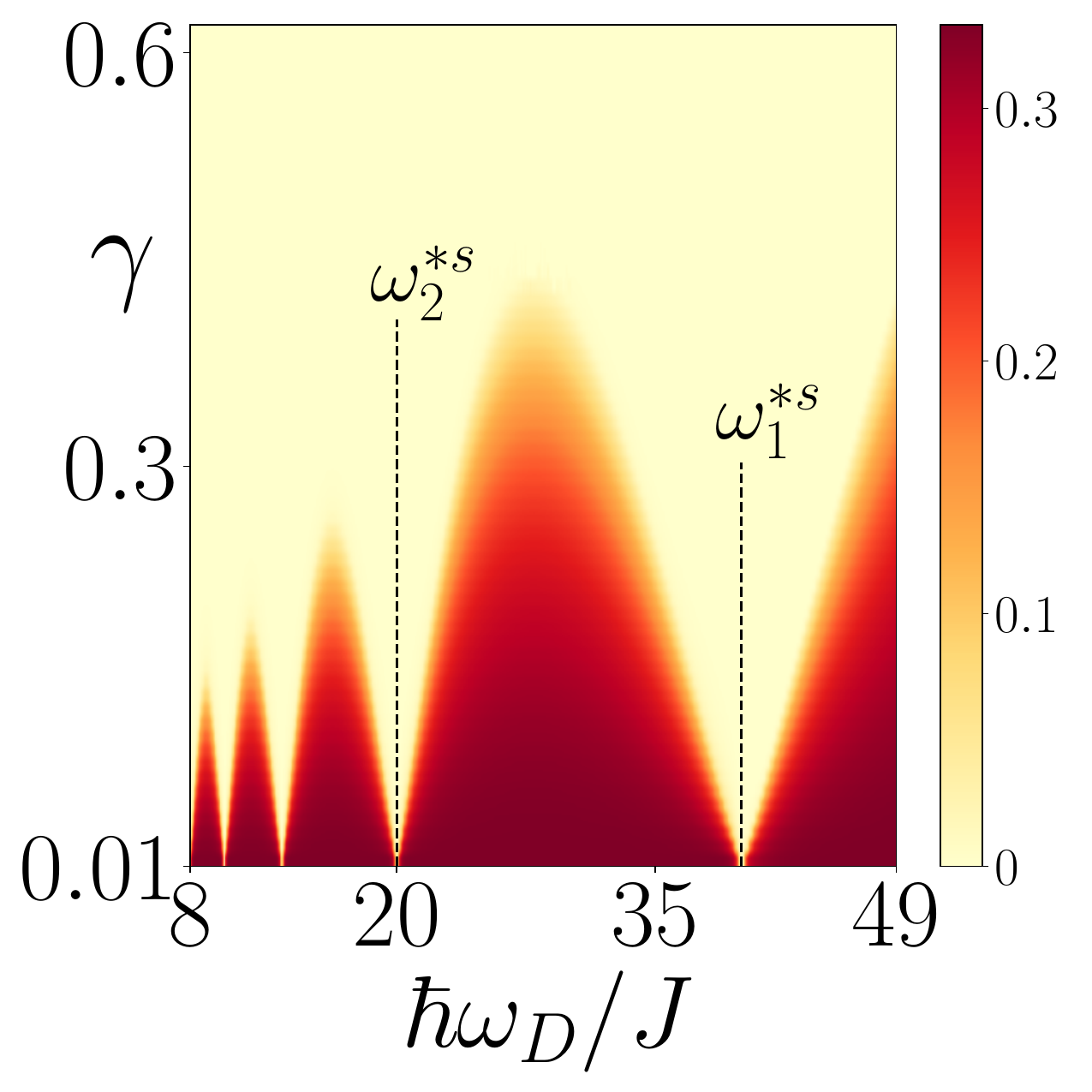}}
\rotatebox{0}{\includegraphics*[width= 0.48 \linewidth]{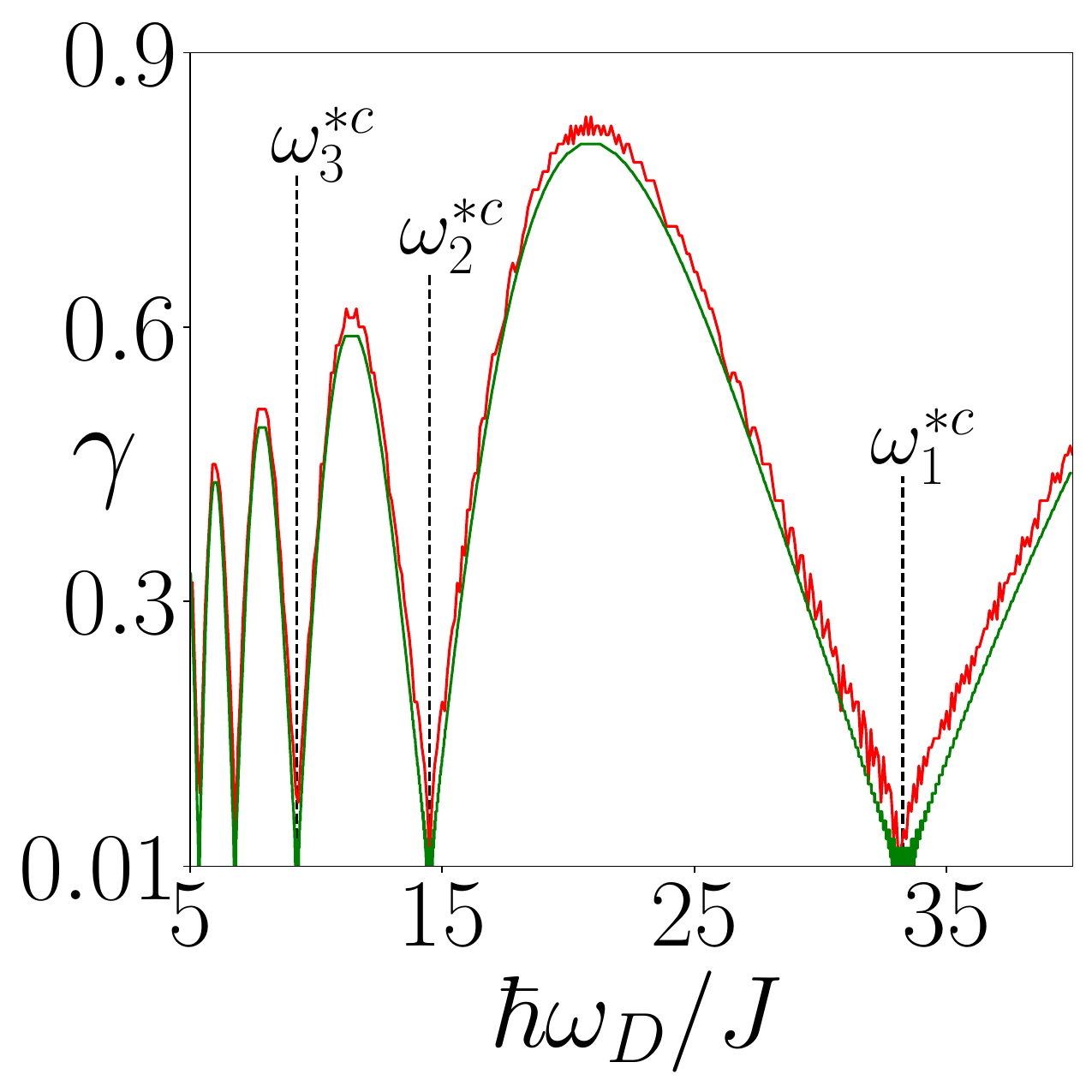}}
\rotatebox{0}{\includegraphics*[width= 0.48 \linewidth]{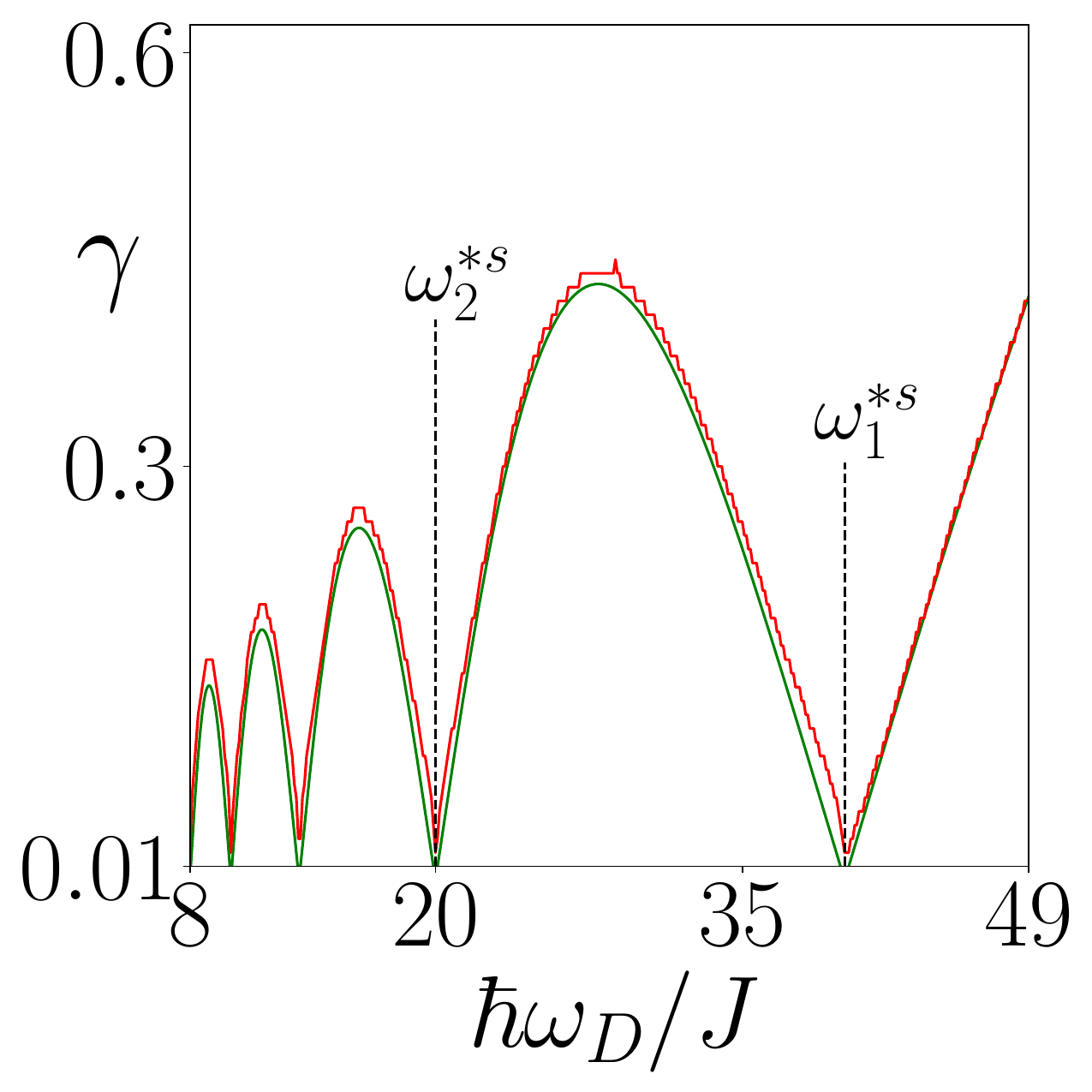}}
\caption{Top left Panel: Plot of $\alpha$, obtained from exact
numerics, as a function of $\omega_D$ and $\gamma$ showing the
non-monotonic nature of $\gamma_c$ and the positions of the special
frequencies $\omega_m^{\ast c}$ for continuous wave drive protocol.
Top right Panel: Same as top left panel obtained using square pulse
drive protocol. Bottom left panel: This figure shows the phase
boundary ($\alpha<0.001$) obtained using exact numerics (red line)
and analytical result (green line which corresponds to $\delta=0$)
for continuous wave drive. Bottom right panel: Same as bottom left
panel obtained for square pulse drive protocol with $\alpha < 0.01$.
For all plots $h_0=20 J$, $h_1=0.1 J$ and $L \le 1000$. See text for
details. \label{fig5}}
\end{figure}
The results obtained using the above procedure for high drive
amplitude and frequency region are presented and compared with their
analytical counterparts (computed using perturbative
Floquet Hamiltonian obtained in Sec.\ \ref{anres} and Apps.\
\ref{appa} and \ref{appb}) in Sec.\ \ref{phd}. This is followed by
Sec.\ \ref{lient} where we present our results for the low and
intermediate drive frequency regime. Most of our
numerical results shall be carried out using continuous drive
protocol; however, we shall also present the phase diagram
corresponding to the square-pulse protocol.

\subsection{Phase diagram in the high drive amplitude regime}
\label{phd}

In the high drive amplitude and frequency regime and away from the
special frequencies $\omega_m^{\ast}$, the half-chain entanglement
entropy $S_{\ell=L/2} \equiv S_{L/2}$ shows two distinct behaviors
as shown in the left panel Fig.\ \ref{fig3} as a function of $L$ for
the continuous protocol. For low $\gamma \simeq 0.1$, $S_{L/2}
\simeq \alpha \ln L $; the coefficient $\alpha \sim 1/3$ in the
regime of high frequency. We note that this result coincides with
the behavior of $S_{\ell}$ (Eq.\ \ref{entrexp}) computed in Sec\
\ref{anres} in the high-frequency and low $\gamma$ regime where
$\delta \to 1$. In contrast, for $\gamma=2$, $S_{L/2}$ displays area
law (inset of left panel of Fig.\ \ref{fig3}) and is thus
independent of $L$. We note that this behavior is qualitatively
different from that of $S_{L/2}$ at $\omega_D=\omega_1^{\ast c}$,
where it shows area law behavior for almost all $\gamma$ as shown in
the right panel of Fig.\ \ref{fig3}.

For $\omega_D \ne \omega_m^{\ast}$, a transition between these two
behaviors occur at a critical $\gamma_c$, as shown in the left panel
of Fig.\ \ref{fig4}, where $\alpha$ is plotted as a function of
$\gamma$. The analytic expression of $S_{\ell}$ predicts $\alpha =
\delta/3$ leading to vanishing of $\alpha$ at $\gamma^{\ast} \simeq
|\tilde{h}_1 f(T)|$ ($\delta=0$). This prediction seems to match the
numerical result qualitatively as also shown in the left panel of Fig.\
\ref{fig4}. The right panel of Fig.\ref{fig4} shows similar results
for the square pulse drive protocol. Note that we do not expect a
quantitative agreement here since the analytic computation of
$S_{\ell}$ is expected to be accurate only for $\delta \sim 1$.

The phase diagram demonstrating this entanglement transition as a
function of the drive frequency $\omega_D$ and the measurement rate
$\gamma$ is given in Fig.\ref{fig5}. The top left panel of Fig.\
\ref{fig5} plots $\alpha$ as a function of $\gamma$ and $\omega$
obtained from fitting the data for $L \le 1000$; this data is
obtained using trotter decomposition of the evolution operator for
the continuous drive protocol. The region where $\alpha \simeq 0$
marks the boundary for entanglement transition as shown in the
bottom left panel of Fig. \ref{fig5}. The numerical cutoff for
obtaining this boundary is set to be $\alpha<0.001$. A similar phase
diagram and boundary for the square pulse drive protocol is shown in
the right panels of Fig. \ref{fig5}. The red lines in the bottom
panels indicate the numerical phase boundaries. The plots clearly
demonstrate the non-monotonic dependence of the phase boundary of
the entanglement transition on the drive frequency. The green lines
in the bottom panels of Fig.\ \ref{fig5} indicate the curves
$\delta=0$ for the continuous (bottom left) and the square pulse
(bottom right) drive protocols. It is evident that in this regime
the analytical result matches exact numerics quite well.

This behavior of $S_{\ell}$ can also be qualitatively understood
from the first order Floquet Hamiltonian as follows. The energy
eigenvalues corresponding to the first order Floquet Hamiltonian
(Eq.\ \ref{flham}) is given by $E_k^{\pm} = \pm E_k = \pm 2
\sqrt{(h_1 -\cos k + i\gamma/2)^2 + (f^{(c,s)}(T)  \sin k)^2}$. For
small $\gamma$ and $f^{c(s)}(T) \ne 0$, $\gamma^2 < 4[(h_1 - \cos
k)^2 + (f^{(c,s)}(T)\sin k)^2]$ and one has $\Gamma_k = {\rm Im}
[E_k] \sim \gamma (h_1 -\cos k)$. Thus $\Gamma_k$ changes sign
around $k^{\ast} \simeq \arccos(h_1)$ leading to jumps in
$\text{Sgn}\left[\Gamma_k\right]$, while
$\epsilon_k=\text{Re}\left[E_k\right]$ varies smoothly having the
same signature over the full BZ. This gives rise to jumps in
$\Pi_x(k)$, as can be seen from top left panel of Fig.\ \ref{fig1}.

In contrast for $\gamma^2 \gg 4[ (h_1 -\cos k)^2 + (f^{(c,s)}(T)
\sin k)^2]$, one has $\Gamma_k \sim \gamma$. In this regime,
$\Gamma_k$ does not change sign, while $\epsilon_k$ smoothly changes
sign across $k\sim\pm k^\ast$. Hence no jumps occur in $\Pi_x(k)$
for any $k$ as can be seen from the top left panel of Fig.\
\ref{fig2}. Since the coefficient $\alpha$ of the $\ln \ell$ term in
$S_{\ell}$ receives contribution from the jump singularities in the
correlation function, it vanishes for large $\gamma$ leading to area
law behavior. In contrast for small $\gamma$, the jumps persist and
$\alpha$ remain finite leading to a finite $\ln \ell$ term in
$S_{\ell}$. This allows one to expect two different phases separated
by a transition at $\gamma=\gamma_c$ for which $\alpha(\gamma_c)=0$.
We also note when $f^{c(s)}(T)$ vanish $\Gamma_k \sim \gamma$ for
almost all $\gamma$ while $\epsilon_k$ smoothly changes sign across
$k\sim\pm k^\ast$. The correlation functions therefore remain smooth
for any $\gamma$. This leads to an area-law behavior for almost all
$\gamma$ and $\gamma_c \to 0$.

Thus we find that in the high drive amplitude and frequency regime
it is possible to traverse between different phases (for which
$S_{\ell} \sim \ln \ell$ and $S_\ell \sim {\rm constant}$) through
the entanglement transition by tuning the drive frequency. In
addition, the approximate emergent symmetry at special drive
frequencies at $\omega_m^{\ast}$ shapes the nature of these
transition; near these frequencies, $\gamma^{\ast} \to 0$ and the
phase featuring area law behavior of $S_{\ell}$ dominates. These
features distinguish the behavior of $S_{\ell}$ in periodically
driven systems from their quench counterparts studied in Ref.\
\onlinecite{nhdyn5}.

\subsection{Low and intermediate drive frequencies}
\label{lient}

At low and intermediate drive frequencies, the results obtained from
the perturbative Floquet Hamiltonian does not agree with those
obtained from exact numerics. This is pointed out in App.\
\ref{appb} for the square pulse protocol where the exact Floquet
Hamiltonian can be computed analytically. For the continuous
protocol, the exact Floquet Hamiltonian can be computed numerically
and it shows a similar deviation. In both cases, this feature is
reflected from the bottom panels of Figs.\ \ref{fig1} and \ref{fig2}
where the behavior of $\Pi_{xk}$ computed using exact numerics
deviates drastically from that computed using $H_F^{(2)}$.

\begin{figure}
\rotatebox{0}{\includegraphics*[width= 0.48 \linewidth]{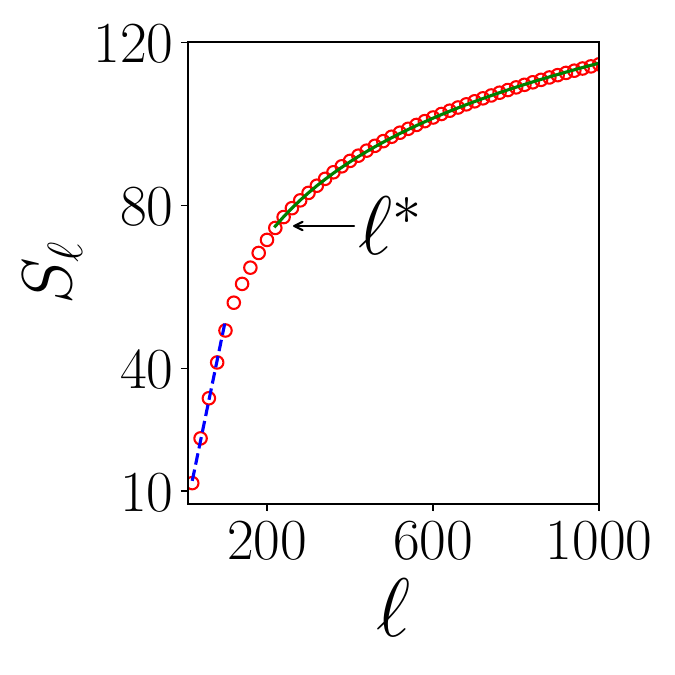}}
\rotatebox{0}{\includegraphics*[width= 0.48 \linewidth]{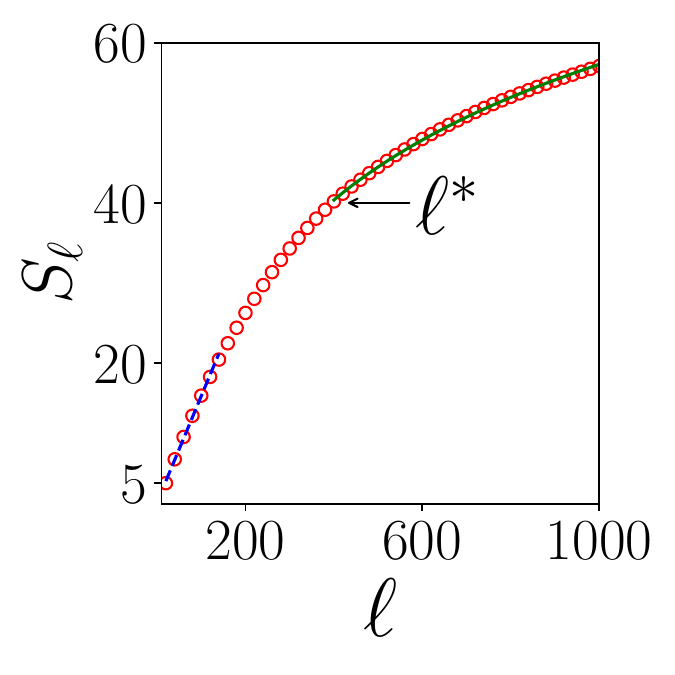}}
\caption{Left Panel: Plot of $S_{\ell}$ as a function of $\ell$ for
$\gamma=0.001 J$ and $\hbar \omega_D/J=0.1$ showing linear behavior at
small $\ell$ for the continuous drive protocol. Right panel: Similar
plot for $\hbar \omega_D/J=0.01$ showing qualitatively similar
behavior. The plots indicate a subsequent crossover to $\ln \ell$
behavior beyond a crossover scale $\ell^{\ast}$ indicated by the
arrows. Note that $\ell^{\ast}$ increases with decreasing drive
frequency. For all plots $h_0=20 J$, $h_1=0.1 J$ . See text for
details. \label{fig6}}
\end{figure}

A numerical computation of $S_{\ell}$ in the low and intermediate
drive frequency regime shows two distinct regimes as  shown in Fig.\
\ref{fig6} for $\hbar \omega_D/J=0.1$ (left panel) and $\hbar
\omega_D/J = 0.01$ (right panel) for $\gamma=0.001 J$. For both of these
drive frequencies, $S_{\ell} \sim \ell$ for $\ell \le \ell^{\ast}$;
in contrast, for $\ell > \ell^{\ast}$, we find $S_{\ell} \sim \ln
\ell$. We note that the emergence of such a volume law for
$S_{\ell}$ does not contradict Szeg\H{o}'s theorem which is valid
for asymptotically large $\ell$. Below, we investigate the origin of
the crossover length scale $\ell^{\ast}$ in the low and intermediate
drive frequency regime.

To this end, we first note that in the low-frequency regime, the
steady state corresponds to an eigenstate of the Floquet Hamiltonian
which corresponds to $\Gamma_k>0$ for each $k$. This situation is to
be contrasted with Hermitian Ising model where it can be a
superposition of both Floquet eiegnvalues at every $k$ \cite{dt2}.
Thus it turns out that if the Floquet Hamiltonian develops long
correlation length in real space in a drive frequency regime, one
would naturally expect a volume-law entanglement for the steady
state when subsystem size is smaller than the correlation length; in
this regime, the effective Hamiltonian describing the subsystem
becomes long-ranged \cite{dt2}. To check if this is indeed the case
here, we resort to the square pulse protocol. Using Eqs.\
\ref{nteq}, \ref{unitdef}, and \ref{flhamsq}, we can express the
Floquet Hamiltonian in real space (up to an additive constant)
\begin{eqnarray}
H'_F  &=& \sum_{j_1 j_2} \left(c_{j_1}^{\dagger} A_{j_1-j_2} c_{j_2}
+
c_{j_1} B_{j_1-j_2} c_{j_2} +{\rm h.c.} \right) \nonumber\\
A_{j_1-j_2} &=& \int_{-\pi}^{\pi} \frac{dk}{2 \pi} e^{-i k
(j_1-j_2)} \hbar \alpha_k n_{zk} \nonumber\\
B_{j_1-j_2} &=& \int_{-\pi}^{\pi} \frac{dk}{2 \pi} e^{i\pi/2} e^{-i
k (j_1-j_2)} \hbar \alpha_k (n_{xk} - i n_{yk}) \label{rspace}
\end{eqnarray}

For the continuous protocol, an analogous expression for
$A_{j_1-j_2}$ and $B_{j_1-j_2}$ exists; however $\alpha_k$ and $\vec
n_k$ need to be obtained numerically. Below we discuss numerical
computation of $A_{j_1-j_2}$; the results obtained for $B_{j_1-j_2}$
are similar and shall not be discussed separately.

\begin{figure}
\rotatebox{0}{\includegraphics*[width= 0.98 \linewidth]{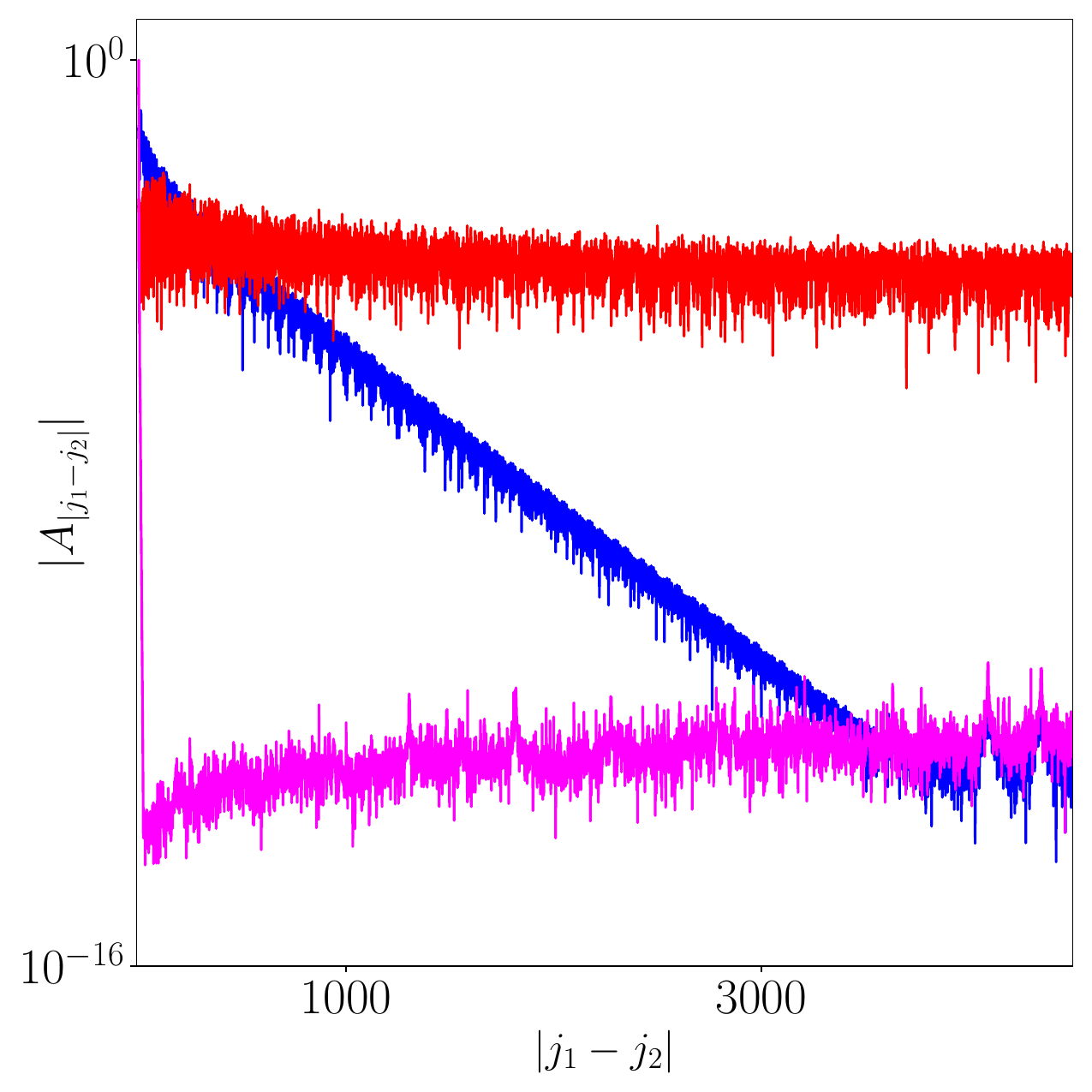}}
\caption{Plot of $A_j$ as a function of $j$ for $\hbar
\omega_D/J=9$ (magenta curve), $1$ (blue curve) and $0.1$ (red
curve) for the square pulse protocol. For all plots $h_0=20 J$,
$h_1=0.1 J$, $\gamma=0.001J$, and $L=10000$. See text for details. \label{fig7}}
\end{figure}

A numerical computation of $A_{j_1-j_2}$ indicates $|A_{j_1-j_2}|
\sim \exp[-|j_1-j_2-1|/\ell^{\ast}(\omega_D)]$ as shown in Fig.\
\ref{fig7} for the square-pulse drive protocol with $\gamma=0.001 J$
and $\hbar \omega_D/J=9,1.0 \,{\rm and}\,0.1$. The length scale
$\ell^{\ast}$, which controls the effective short- or long-range
nature of the Floquet Hamiltonian, can be obtained by standard
fitting procedure.

As can be clearly seen from Fig.\ \ref{fig7}, $\ell^\ast$ increases with
decreasing drive frequency. At high and intermediate drive frequencies,
$\ell^{\ast}(\omega_D) \to 0$ (magneta curve in Fig.\ \ref{fig7})
signifying that the Floquet Hamiltonian is well described by a
nearest-neighbor fermion hopping model. With decreasing
$\omega_D$, $\ell^{\ast}$ increases; this
indicates when the subsystem size $\ell <\ell^{\ast}$, $S_{\ell}$ in
the steady state mimics features of a long-range model. This leads
to a volume-law behavior. In contrast, for $\ell \ge
\ell^{\ast}(\omega_D)$, the entanglement is still due to an
effectively short-range model and we get the expected logarithmic
behavior. The crossover between these two regimes occur around
$\ell=\ell^{\ast}(\omega_D)$ as can be seen from Fig.\ \ref{fig6}.

For large $\gamma$, $\ell^{\ast}$ always remain small. This can be
qualitatively understood from the fact that for $\gamma \gg 1$,
$\alpha_k n_{zk} \sim \gamma + {\rm O}(1)$ and $ \alpha_k n_{x k}
\sim {\rm O}(1)$ and $\alpha_k n_{y k} \sim {\rm O}(1)$. This
indicates $\ell^{\ast} \to 0$ even in the low-drive frequency limit
and one obtains area law behavior of $S_{\ell}$ for all $\ell$.

\section{Discussion}
\label{diss}

In this work, we have studied entanglement transitions in a
periodically driven integrable non-Hermitian Ising model. The
non-hermiticity of the model arises from the presence of a complex
transverse field; such a non-hermitian model can originate from an
Ising chain subjected to measurements in the so-called no-click
limit \cite{dalibard1,nhdyn5}. Our analysis is based on a
Jordan-Wigner transformation which maps such a driven chain to a
system of spinless fermions; this allows to conclude that such a
study is applicable to several other spin systems such as the
non-Hermitian XY model and the Kitaev chain which can be reduced to
the same free fermion form by a Jordan-Wigner transformation
\cite{tb1}.

Our analysis presents a detailed phase diagram for entanglement
transition in such driven system. For high drive frequency regime,
we find two phases. In the first phase which occurs when the
imaginary part of the transverse field, $\gamma$, is small, the
steady state entanglement shows a logarithmic dependence on the
sub-system size. We provide an explicit analytic expression for the
coefficient of the $\ln \ell$ term, $\alpha$, in the small $\gamma$
limit and show that it can be tuned using drive frequency. In
particular, we find that $\alpha \to 0$ for almost all $\gamma$ at
some special drive frequencies; this is a consequence of an
approximate emergent symmetry of the driven model. In contrast, for
large $\gamma$, the entanglement exhibits an area-law behavior. We
chart out the phase boundary between these phases as a function of
$\gamma$ and $\omega_D$. Our analytic results based on contribution
of the Fisher-Hartwig singularities to $S_{\ell}$ indicate
$\alpha=\delta/3$. This result matches exact numerics in the high
frequency regime quite well and shows that the entanglement
transition boundary is well-approximated by the relation $\delta=0$.
We note that the tuning of $\alpha$ using the drive frequency and
the presence of special frequencies where $\alpha \to 0$ due to an
approximate emergent symmetry has not been presented earlier.

We also numerically study the entanglement of the driven system at
low and intermediate drive frequencies where analytic results are
difficult to obtain. In this regime, we identify a length scale
$\ell^{\ast}(\omega_D)$ which can be identified as the correlation
length of the driven Ising chain. For subsystem size $\ell \le
\ell^{\ast}$, the driven chain sees an effective long-range Floquet
Hamiltonian for small $\gamma$. Consequently, it exhibits volume law
entanglement. As one increase $\ell$, $S_{\ell}$ crosses over to a
$\ln \ell$ behavior when $\ell > \ell^{\ast}$. The value of
$\ell^{\ast}$ increase with decreasing $\omega_D$. In contrast, for
large $\gamma$, the driven chain always has short range correlation
in the steady state; consequently, $S_{\ell}$ always exhibits
area-law. We note that the emergence of such a length scale for
driven non-hermitian chains has not been pointed out in the
literature before.

There are several possible extensions to our work. The first of
these constitutes an exact calculation of $S_{\ell}$ using Fredholm
techniques for non-hermitian Ising systems; such a calculation would
be an interesting application of the technique in the domain of
non-hermitian quantum systems. The second is the possibility to
study the entanglement of such a system away from the no-click
limit; this requires analysis of the Ising chain coupled to the
detector in its full generality. It will be useful to check if the
entanglement transition survives in this case; this question is of
relevance to possible experiments and is of particular importance in
the low $\omega_D$ and low $\gamma$ limit where the system takes a
long time to reach the steady state. We leave these issues for
future work.

In conclusion, we have studied entanglement transition in a driven
non-Hermitian Ising chain and charted out the corresponding phase
boundary. We have provided analytical results for $S_{\ell}$ in the
high drive frequency regime and discussed its relation to an
emergent approximate symmetry of the driven chain. In the low and
intermediate drive frequency regime, we have identified a
correlation length scale which shapes the behavior of the
entanglement at low $\gamma$. We expect our results to be applicable
to other, similar, non-hermitian integrable models such as the XY
and Kitaev chains.

\section{Acknowledgement}

The authors thank M. Schiro, A. Silva and J. De Nardis for discussions.
KS thanks DST, India for support through SERB project
JCB/2021/000030.

\appendix

\section{Floquet Hamiltonian}
\label{appa}

In this appendix, we sketch the derivation of the Floquet
Hamiltonian starting from Eq.\ \ref{fermham} of the main text. We
first concentrate on the high drive amplitude regime where $h_0 \gg
h_1, J ,\gamma$. In this regime one can write $H= H_0(t) +H_1$ where
\begin{eqnarray}
H_0(t) &=& \sum_k \psi_k^{\dagger} 2h_0(t) \sigma_z \psi_k
 \nonumber\\
H_1 &=& 2 \sum_k \psi_k^{\dagger}[  (h_1 - \cos k + i \gamma/2)
\sigma_z
\nonumber\\
&&  + (\sigma_+ \sin k + {\rm h.c.})] \psi_k \label{hd1}
\end{eqnarray}
We note that $h_0(t)$ depends on the protocol used and can be read
off from Eqs.\ \ref{contprot} and \ref{sqprot} for continuous and
discrete drive protocols respectively.

To obtain the Floquet Hamiltonian from Eq.\ \ref{hd1}, we first
construct the evolution operator corresponding to $H_0(t)$:
$U_0(t,0) \equiv U_0(t)= \exp[-i \int_0^t H_0(t') dt'/\hbar]$. A
straightforward evaluation leads to \cite{tb1}
\begin{eqnarray}
U_0^c(t) &=& \prod_k e^{-2 i h_0 \sin\left(\omega_D t\right) \psi_k^{\dagger}
\sigma_z \psi_k/(\hbar \omega_D)} \label{evcontprot}
\end{eqnarray}
for the continuous protocol and
\begin{eqnarray}
U_0^s(t) &=& \prod_k e^{2 i h_0 t \psi_k^{\dagger} \sigma_z
\psi_k/\hbar} \quad t \le T/2, \nonumber\\
&=& \prod_k e^{2 i h_0 (T-t) \psi_k^{\dagger} \sigma_z \psi_k/\hbar}
\quad t > T/2 . \label{evsqprot}
\end{eqnarray}
for the square pulse drive protocol. Note that $U_0(T)=I$ for both
continuous and square-pulse protocols which indicates that
$H_F^{(0)}=0$.

The first-order contribution to the Floquet Hamiltonian can be
computed perturbatively using standard prescription of FPT
\cite{rev8,tb1}. One obtains
\begin{eqnarray}
U_1^{c(s)}(T) &=& \frac{-i}{\hbar} \int_0^T dt\,
[U_0^{c(s)}(t)]^{\dagger} H_1 U_0^{c(s)}(t)  \label{fofpt}
\end{eqnarray}
This can be evaluated in a straightforward manner following Ref.\
\onlinecite{tb1} and leads to
\begin{eqnarray}
H_F^{(1),c(s)} &=& i \hbar U_1^{c(s)}(T)/T \nonumber\\
&=& \sum_k \psi_k^{\dagger}[
S_{1k} \sigma_z + ( S_{2k} \sigma^+ + {\rm h.c.})]\psi_k \nonumber\\
S_{1k} &=& 2 (h_1- \cos(k)+ i\gamma/2), \label{flham4} \\
S_{2k} &=& 2 f^{c}(T) \sin k, \quad {\rm cosine\, protocol}  \nonumber\\
&=& 2 f^{s}(T) \sin k \;e^{-ih_0T/\hbar}, \quad {\rm square\, pulse\,
protocol} \nonumber
\end{eqnarray}
where
\begin{eqnarray}
f^c(T) &=& J_0\left(4h_0/(\hbar \omega_D) \right) \nonumber\\
f^s(T) &=& \frac{\hbar\sin\left( h_0T/\hbar\right)}{h_0T}
\label{f1ham3}
\end{eqnarray}
The higher order corrections to the Floquet Hamiltonian has been
computed in Ref.\ \onlinecite{tb1}. These terms in $p^{\rm th}$
order of perturbation theory, are typically suppressed by a
$1/\omega_D^{p-1}$ factor compared to $H_F^{(1)}$. For $p=2$, such
terms merely change the form of $S_{1k}$ and $S_{2k}$; they do not
alter the structure of $H_F$. An explicit computation carried out in
Ref.\ \onlinecite{tb1} shows that
\begin{eqnarray}
S_{1k} &=& (\alpha_{1 k} + i \gamma) \quad
S_{2 k} = \Delta_{k} (\alpha_{2 k} + i
\gamma \lambda ) \label{sord}
\end{eqnarray}
where $\alpha_k = 2(h_1 -\cos k)$, $\Delta_k = 2\;\sin k$ and
\begin{eqnarray}
\alpha_{1 k} &=& \alpha_{k} -2 \Delta^2_{k} \sum_{n=0}^{\infty}
\frac{ J_0\left(\frac{4h_0}{\hbar \omega_D}\right)
J_{2n+1}\left(\frac{4h_0}{\hbar
\omega_D}\right)}{(n+1/2) \hbar \omega_D} \nonumber\\
\alpha_{2 k} &=& J_0\left(\frac{4h_0}{\hbar \omega_D}\right) +
\alpha_{k}  \lambda \nonumber\\
\lambda  &=& 2 \sum_{n=0}^{\infty}
\frac{J_{2n+1}\left(\frac{4h_0}{\hbar \omega_D}\right)}{(n+1/2)
\hbar \omega_D} \label{cexp}
\end{eqnarray}
These expressions for $S_{1k}$ and $S_{2 k}$ are used for computing second-order
perturbative results for the correlation matrix in the main text.

\section{Exact $H_F$ for the square pulse protocol}
\label{appb}

The evolution operator for the square-pulse drive protocol given by
Eq.\ \ref{sqprot} of the main text is
\begin{eqnarray}
U_{k} &=& \prod_k  e^{-iH_k^+T/(2\hbar)} e^{-i H_k^- T/(2\hbar)} =
\prod_k e^{-i H_{F k}T/\hbar}
\nonumber\\
H_k^{\pm} &=& \theta_k^{\pm} \left( r_{zk}^{\pm} \tau_z + r_{x
k}^{\pm} \tau_x \right) \label{sqham1}
\end{eqnarray}
where
\begin{eqnarray}
\theta_k^{\pm} &=& 2 \sqrt{\left(h_1 \pm h_0 +i \gamma/2- \cos k
\right)^2 + \sin^2 k} \nonumber\\
r_{z k}^{\pm} &=& \frac{2(h_1 \pm h_0 +i\gamma/2- \cos
k)}{\theta_k^{\pm}}, \,\, r_{xk}^{\pm} = \frac{2 \sin
k}{\theta_k^{\pm}}. \label{nteq}
\end{eqnarray}
and we have dropped the subscript $s$ used in the main text
indicating square protocol here and for the rest of this section.

The expression of $H_{Fk}$ can be obtained in a straightforward
manner. To express this concisely, we define a set of new unit
vectors $\vec n_k = (n_{x k}, n_{y k}, n_{z k})^T$ which are related
to $r_{x k}^{\pm}$ and $r_{z k}^{\pm}$ by
\begin{widetext}
\begin{eqnarray}
n_{z k}\sin(\alpha_k T) &=&  r_{z k}^{-} \cos (\theta_k^+
T/(2\hbar)) \sin (\theta_k^- T/(2 \hbar)) +
r_{zk}^{+}\cos (\theta_k^- T/(2 \hbar)) \sin (\theta_k^+ T/(2\hbar)) \nonumber\\
n_{y k} \sin(\alpha_k T) &=& \sin (\theta_k^- T/(2\hbar))\sin
(\theta_k^+ T/(2 \hbar)) (
r_{xk}^{-} r_{z k}^{+} - r_{z k}^{-} r_{x k}^{+} ) \nonumber\\
n_{x k} \sin(\alpha_k T) &=& \cos (\theta_k^+ T/(2 \hbar)) \sin
(\theta_k^- T/(2 \hbar)) r_{x k}^{-}
+\cos (\theta_k^- T/(2\hbar)) \sin (\theta_k^+ T/(2 \hbar)) r_{x k}^{+} \nonumber\\
\cos(\alpha_k T) &=&  \cos (\theta_k^- T/(2 \hbar))\cos (\theta_k^+
T/(2 \hbar)) - \left( r_{z k}^{-} r_{z k}^{+} + r_{x k}^{-}r_{x
k}^{+} \right) \sin (\theta_k^- T/(2 \hbar))\sin (\theta_k^+ T/(2
\hbar)) \label{unitdef}
\end{eqnarray}
\end{widetext}
A few lines of straightforward manipulations show that the Floquet
Hamiltonian can be expressed in terms of these unit vectors and
$\alpha_k$ as
\begin{eqnarray}
H_{F k} &=& \hbar \alpha_k \vec {n}_{k} \cdot \vec{\sigma} \label{flhamsq}
\end{eqnarray}
In the large drive amplitude regime where $h_0 \gg h_1, J $, $n_{yk}
\to 2\sin k(1-\cos(2h_0 T/\hbar))/(2h_0T\alpha_k)$, $n_{z k} \to
2(h_1-\cos k +i \gamma/2)/(\alpha_k \hbar)$ and $n_{x k} \to 2\sin
k\sin(2h_0T/\hbar)/(2h_0T\alpha_k)$. Thus we recover Eq.\
\ref{flham} of the main text.

The Floquet eigenvalues and eigenvectors can be obtained from Eq.\
\ref{flhamsq} via solution of $H_{Fk} \Psi_k^{\pm} = E_k^{\pm}
\Psi_k^{\pm}$. This leads to
\begin{eqnarray}
E_k^{\pm} &=& \pm \hbar \alpha_k, \quad  \psi_k =
\begin{pmatrix}
         u_k^{\pm} \\ v_k^{\pm}
    \end{pmatrix} \label{eigenres} \\
  \begin{pmatrix}
            u_k^{\pm} \\ v_k^{\pm}
    \end{pmatrix} &=& \begin{pmatrix}
        n_{zk} \pm 1 \\ n_{xk} + in_{yk}
    \end{pmatrix} \frac{1}{\sqrt{\left |n_{zk}\pm1\right |^2+ \left|n_{xk} + i n_{yk}\right|^2
    }} \nonumber
\end{eqnarray}

The wavefunction after $n$ cycles of the drive can therefore be
written, in terms of $u_k^{\pm}$ and $v_k^{\pm}$ as given in Eq.\
\ref{wavf1} of the main text. Here we note that the steady state
wavefunction corresponds to
\begin{eqnarray}
|\psi_k^{{\rm steady}}\rangle = \frac{u_k^{+(-)} + v_k^{+(-)}
c_k^{\dagger} c_{-k}^{\dagger}|0\rangle}{\sqrt{|u_k^{+(-)}|^2+
|v_k^{+(-)}|^2}} \label{wavst}
\end{eqnarray}
if ${\rm Im}[E_k]= \Gamma_k > 0(<0)$. Thus the steady state changes
with sign change of $\Gamma_k$. Combining these results, we obtain
the final form of the steady states used in the main text
\begin{eqnarray}
|\psi^{\text{steady}}\rangle &=& \prod_{k>0} \left( \;
u_k^{\text{steady}}|0\rangle + v_k^{\text{steady}}\;
\hat{c}_k^\dagger \hat{c}_{-k}^\dagger|0\rangle \right) \nonumber\\
u_k^{\text{steady}} &=& \frac{S_{z k} + E_k
\;\text{sgn}(\Gamma_k)}{C_k} \quad v_k^{\text{steady}} = \frac{S_{x
k} + iS_{y k} }{C_k} \nonumber\\
S_{j k} &=& \alpha_k n_{jk} \hbar,\; C_k =
\sqrt{|u_k^{\text{steady}}|^2+|v_k^{\text{steady}}|^2}
\label{steadyexp}
\end{eqnarray}

A similar procedure can be carried out for obtaining the steady
state corresponding to the continuous drive protocol. However, for
such a protocol, analytic expressions for Floquet eigenstates and
eigenvalues can not be obtained; in the main text, we have obtained
these quantities numerically and used them to obtain the exact
steady state wavefunctions.

\section{Szeg\H{o}'s theorem and the absence of volume law scaling}
\label{appc}

In this appendix, we show the absence of volume law scaling for
large $\ell$ using Szeg\H{o}'s strong limit theorem \cite{sth1}. We note that this
analysis ignores the jump singularities of $\Pi_{x k}$ that are
separately analyzed in Sec.\ \ref{anres} of the main text.

We begin by noting that $\hat{\Pi}_k= (\Pi_{xk}, \Pi_{yk}, \Pi_{z
k})$ satisfies
\begin{eqnarray}
\left|\left| \; \hat{\Pi}_k \; \right|\right|^2 &=& 1
\nonumber\\
 \text{Det}\left(\hat{\Pi}_k\right) &=&
-\left|\left| \; \hat{\Pi}_k \; \right|\right|^2 = -1
\label{eq:symbol_det}
\end{eqnarray}
This property follows from the normalization of the driven
wavefunction as discussed in Sec.\ \ref{anres}. We also note that
the Fourier transform of $\hat{\Pi}(k)$, defined as,
\begin{eqnarray}
\Pi_{\ell} &=& \int_{-\pi}^{+\pi} \frac{dk}{2\pi} \; e^{ -i k \ell}
\; \hat \Pi(k) \cdot \hat \sigma
 \label{pirel}
\end{eqnarray}
serves as the elements of the correlation matrix $B_{\ell}$. These
elements are matrix valued and in terms of them one obtains
\begin{eqnarray}
G_{\ell} &=& \left ( \begin{array}{cccc} \Pi_0 & \Pi_{-1} & ... & \Pi_{1-\ell} \\
\Pi_1 & \Pi_{0} & ... & \Pi_{2-\ell} \\
... & ... & ... & ... \\ \Pi_{\ell-1} & \Pi_{\ell-2} & ... & \Pi_{0}
\end{array} \right), \label{eq:corr-mat}
\end{eqnarray}
Thus the correlation functions lead to a block Toeplitz matrix.

For later convenience, we next introduce the function $\displaystyle
\mathcal{D}_{\ell}(\lambda)$ given by
\begin{equation}
\mathcal{D}_{\ell}(\lambda) \equiv \lambda I - G_{\ell} \label{ddef}
\end{equation}
where $I$ denotes the identity matrix. In addition, following
standard procedure in the literature \cite{fwref1,fwref2}, we also
define the function $e(x,\lambda)$
\begin{equation}
e(x,\lambda) \equiv -\frac{x+\lambda}{2}
\ln{\frac{x+\lambda}{2}}-\frac{x-\lambda}{2}
\ln{\frac{x-\lambda}{2}} \label{ldef}
\end{equation}
It is well known that in terms of this function the von-Neumann
entanglement entropy can be written as $S_{\ell} =
\sum_{m=-\ell}^{\ell} e(1,\nu_m)$ where $\nu_m$ denote the
eigenvalues of $G_{\ell}$.

Next, we note that from Eq.~\eqref{ddef} we can obtain
\begin{equation}
    \text{Det}\left( \mathcal{D}_{\ell}(\lambda) \right) = \prod_{m=1}^{\ell}
    (\lambda^2-\nu_m^2) \label{dlabel}
\end{equation}
where we have used the property $\nu_{m}= -\nu_{-m}$. Using Cauchy's
residue theorem and Eq.\ \ref{dlabel} we can therefore express
$S_{\ell}$ as a complex integral as
\begin{equation}
\mathcal{S}_{\ell}=\frac{1}{4\pi i} \oint_{\mathcal{C}} \; d\lambda
\: e(1,\lambda) \frac{d}{d\lambda} \ln\left[ \text{Det}\left(
\mathcal{D}_{\ell}(\lambda) \right) \right]
\end{equation}
where $\displaystyle\mathcal{C}$ is the contour surrounding the
zeros of $\displaystyle\text{Det}\left[ \mathcal{D}_{\ell}(\lambda)
\right]$.

To obtain the linear term in $S_{\ell}$, we first define
\begin{equation}
\mathcal{A}_k = \lambda I- \hat{\Pi}_k
\end{equation}
and use Szeg\H{o}'s strong limit theorem. In the leading order this
theorem allows one to write \cite{sth1}
\begin{equation}
\mathcal{S}_{\infty}(\ell) \approxeq \frac{\ell}{8\pi^2 i}\;
\oint_{\mathcal{C}} \; d\lambda \: e(1,\lambda) \; \int_{0}^{2\pi}
\; dk \: \frac{d}{d\lambda} \ln\left[ \text{Det}\left( \mathcal{A}_k
\right) \right] \label{eq:entropy_formula_steady_state}
\end{equation}
Using Eq.\ \ref{eq:symbol_det}, we find that ${\rm Det} ({\mathcal
A}_k)= \lambda^2-1$. Using this,
Eq.~\eqref{eq:entropy_formula_steady_state} gives us
\begin{widetext}
\begin{eqnarray}
\mathcal{S}_{\text{steady}}(\ell) & \approxeq & \frac{\ell}{8\pi^2 i} \oint_{\mathcal{C}}\; d\lambda \: e(1,\lambda) \int_{0}^{2\pi}\; dk \: \frac{d}{d\lambda}\left[ \ln(\lambda-1) +\ln(\lambda+1) \right] \nonumber\\
&=& \frac{\ell}{8\pi^2 i} \oint_{\mathcal{C}}\; d\lambda \: e(1,\lambda) \int_{0}^{2\pi}\; dk \: \left[ \frac{1}{\lambda-1} + \frac{1}{\lambda+1} \right]  \nonumber\\
&=& \frac{\ell}{4\pi i} \oint_{\mathcal{C}}\; d\lambda \: \left[
\frac{e(1,\lambda)}{\lambda-1} + \frac{e(1,\lambda)}{\lambda+1}
\right]=0
\end{eqnarray}
\end{widetext}
where the last line follows from the presence of simple poles at
$\lambda=\pm 1$.

Thus we arrive at the conclusion that the coefficient of the linear
term vanishes. This indicates that $S_{\ell}$ can show either
logarithmic dependence on $\ell$ or lead to an area law behavior
($S_{\ell}$ being a constant) for large $\ell$. We show in the main
text using a Fisher-Hartwig type analysis that the first behavior
follows from the jump singularities of $\Pi_{x k}$ at small $\gamma$
while the latter behavior is seen at large $\gamma$ where such jumps
are absent.

\section{Fisher-Hartwig singularities of $\Pi_{x k}$}
\label{appd}

\begin{figure}\rotatebox{0}{\includegraphics*[width= 0.98
\linewidth]{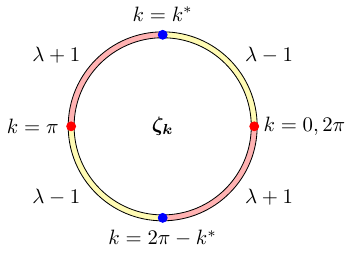}} \caption{Schematic representation of the
jump singularities which separates the Brillouin zone into different
regions. The solution of $\zeta_k$ is obtained by equating it to
$\lambda\pm 1$ in these regions as schematically shown. See text for
details. \label{figappd1}}
\end{figure}

In this section, we sketch the derivation of ${\mathcal D}_{\ell}$
starting from the expression of $\Pi_{xk}^{(1)}$ (Eq.\
\ref{pikform}) in the main text. We are going to first sketch this
derivation for $\delta=1$ where the norm of $\Pi_{x k}$ is unity. In
the rest of the appendix, we shall shift the Brillouin zone from $0
\le k \le 2 \pi$ instead of $-\pi \le k \le \pi$. The singularities
therefore occur at $k=k^{\ast}$ and $2\pi-k^{\ast}$. This allows us
to write
\begin{eqnarray}
\tilde\Pi_k^{(1)} &=& g(k) \; \sigma^x \quad 0 \leq k < 2\pi
\nonumber\\
\zeta_k &=& \lambda - g(k) \label{starteq}
\end{eqnarray}
where $g(k)=\pm 1$. We perform the calculations in the standard form
by working on the unit circle $S^1$: $z= e^{ik}$. The generator,
$\zeta(z)$ contains Fisher-Hartwig jump singularities as shown
schematically in Fig.\ \ref{figappd1}; hence one can apply the
Fisher-Hartwig conjecture to find the asymptotic value of the
Toeplitz determinant $\mathcal{D}_{\ell}(\lambda)$ which in turn
will allow us to obtain the asymptotic analytic form of $S_{\ell}$.

To this end, we express the generator $\zeta(z)$ in the
Fisher-Hartwig form. We denote $z_j = e^{i k_j}$, where $k_j = 0,
k^\ast,\pi,(2\pi-k^\ast)$, leading to $z_j = 1, z^{\ast}, -1,
1/z^{\ast}$, where $z^{\ast}= \exp[i k^{\ast}]$, to be the positions
of the jump discontinuities. We note that the discontinuities at $z
= \pm 1$ are spurious as explained in the main text. We divide the
unit circle $S^1$ in four regions as shown in Fig.\ \ref{figappd1}.
In each region, the generator $\zeta(z)$ has a constant value
$\lambda \pm 1$.

We also define $g_{z_j\beta_j}(z)\text{ for }j=1,2,3$ that changes
across the discontinuity points and $g_{z_0\beta_0}$ which remains
the same over the entire unit circle; this allows us to choose
$g_{z_0 \beta_0}= \exp[- i \pi \beta_0]$. The other
$g_{z_j\beta_j}(z)$ has the following structure.
\begin{eqnarray}
g_{z_j,\beta_j}(z) & = &  e^{-i\pi\beta_j} \quad 0<k<k_j \nonumber \\
& = & e^{i\pi\beta_j} \quad k_j<k<2\pi
\label{eq:FH-form}
\end{eqnarray}
In terms of this function the generator takes the following form
\begin{eqnarray}
\zeta(z) &=& e^{V(z)} \left( \frac{z}{z_0} \right)^{\beta_0} \left(
\frac{z}{z_1} \right)^{\beta_1} \left( \frac{z}{z_2}
\right)^{\beta_2} \left( \frac{z}{z_3} \right)^{\beta_3}
\nonumber\\
&& \times e^{-i\pi\beta_0} g_{z_1\beta_1}(z)g_{z_2\beta_2}(z)g_{z_3\beta_3}(z)
\label{eq:FH-form}
\end{eqnarray}

In order to find the functional form of the functions
$g_{z_j\beta_j}(z)$, one needs to solve a set of coupled
equations. It is clear from Fig.\ \ref{figappd1} that the condition
$\zeta(z) = \lambda \pm 1$ necessitates $V(z)=\ln \eta_0(\lambda)$
to be independent of $z$ and
\begin{equation}
\beta_0+\beta_1+\beta_2+\beta_3=0 \label{eq:beta_condition}
\end{equation}
With these conditions the generator takes the following form
\begin{eqnarray}
\zeta(z) &=&  \eta_0(\lambda)
\prod_{j=1}^3 z_j^{\beta_j}
e^{-i\pi\beta_0} g_{z_1\beta_1}(z)g_{z_2\beta_2}(z)g_{z_3\beta_3}(z)
\nonumber\\ \label{zetasol}
\end{eqnarray}

With this form of $\zeta(z)$, we now seek the solution of the
equations
\begin{eqnarray}
\zeta(z) &=& \lambda \pm 1  \label{zetaeq}
\end{eqnarray}
as shown schematically in Fig.\ \ref{figappd1}. The solutions of
these equations are straightforward and yield
\begin{eqnarray}
\beta_0 &=& \beta_2=\frac{1}{2\pi i}\ln\left(
\frac{\lambda+1}{\lambda-1} \right) \nonumber\\
\beta_1 &=& \beta_3=\frac{1}{2\pi i}\ln\left(
\frac{\lambda-1}{\lambda+1} \right) \nonumber\\
\eta_0(\lambda) &=& \sqrt{\lambda^2-1} \label{solparam}
\end{eqnarray}
The above solution above is for $\text{Sgn}\left(\tilde{h}_1
f(T)\right)=1$. Solution for $\text{Sgn}\left(\tilde{h}_1
f(T)\right)=-1$ can be obtained by exchanging $\lambda+1$ and
$\lambda-1$. The final expression of $S_{\ell}$ will remain the same
in both cases.

The form of $\zeta(z)$ (Eq.\ \ref{zetaeq}) obtained allows us to
write the matrix $G_{\ell}= B_{\ell} \otimes \sigma^x$, where
$B_{\ell}$ is given by Eq.\ \ref{corrmat1} in the main text. To this
end, we define
\begin{equation}
\mathcal{D}_{\ell}\left( \lambda \right) = \lambda \mathbb{I}_{2
\ell\times 2\ell} -G_{\ell}
\end{equation}
where $G_{\ell}$ is a $ 2\ell\times 2\ell$ Toeplitz matrix. As all
the non-trivial correlations are embedded within the
$\ell\times\ell$ Toeplitz matrix $B_{\ell}$, we compute the
asymptotics of $\text{Det}[\tilde{\mathcal{D}}_\ell(\lambda)]$ in
order to obtain the relevant information about the entanglement
entropy of the system, where
\begin{equation}
\tilde{\mathcal{D}}_\ell(\lambda) = \lambda
\mathbb{I}_{\ell\times\ell}-B_\ell.
\end{equation}

Following the Fisher-Hartwig conjecture, we can write
\begin{equation}
\text{Det}\left(\tilde{\mathcal{D}}_{\ell}\left(\lambda\right)
\right) = \left( F\left[ \eta(\lambda)\right]\right)^\ell
\ell^{-\sum_{i=0}^{3} \beta^2_i(\lambda)} \tilde{\bar{E}}
\label{dexpapp}
\end{equation}
where $F[\eta(\lambda)]=\eta_0(\lambda)$ and $\tilde{\bar{E}}$ can
be expressed in terms of Barnes G function \cite{fwref1,fwref2}
whose form is not important for our analysis as long as we only
restrict ourselves to the leading order contribution to $S_{\ell}$.
This allows one to write
\begin{eqnarray}
\ln\left( \text{Det}\left( \mathcal{D}_{\ell}\left(\lambda\right)
\right) \right) &=&  2 \ln\left( \text{Det}\left(
\tilde{\mathcal{D}}_{\ell}\left(\lambda\right)
\right) \right) \nonumber\\
&=& 2[\ell \; \ln\left( F\left[\eta(\lambda)\right]
\right) -\sum_{i=1,3} \beta_i^2(\lambda)\ln(\ell)] \nonumber\\
&& + \text{sub-leading corrections} \label{lndexp}
\end{eqnarray}
The first term, which arises from non-singular part of $\zeta_k$,
does not contribute to $S_{\ell}$. This is seen by application of
Szeg\H{o}'s theorem and is detailed in App.\ \ref{appc}. In the rest of
this section, we shall be concerned with the term $\sim \ln \ell$ in
Eq.\ \ref{lndexp}. Note that we have omitted the contribution of
$\beta_0$ and $\beta_2$; this amounts to subtracting out the
contribution from the spurious singularities at $k=0, \pi$.

The above mentioned calculation may be easily repeated with $\delta
\ne 1$. A procedure, exactly similar to the one outlined above,
requires solution of the equations $\zeta_k = \lambda \pm \delta$
and yields
\begin{eqnarray}
\beta_0 &=& \beta_2=\frac{1}{2\pi i}\ln\left(
\frac{\lambda+\delta}{\lambda-\delta} \right) \nonumber\\
\beta_1 &=& \beta_3=\frac{1}{2\pi i}\ln\left(
\frac{\lambda-\delta}{\lambda+\delta} \right) \nonumber\\
\eta_0(\lambda) &=& \sqrt{\lambda^2- \delta^2} \label{solparam2}
\end{eqnarray}
Substituting Eq.\ \ref{solparam2} in Eq.\ \ref{lndexp}, one finally
gets the expression of $\ln \left[ \text{Det}\left({\mathcal
D}_{\ell}\right)\right]$ used in Eq.\ \ref{dexp1} in the main text.

\end{document}